\documentclass[12pt,a4paper]{article}
\usepackage{a4wide}
\usepackage{amsmath}
\usepackage{amssymb}
\usepackage{epsfig}
\usepackage{subfigure}
\usepackage{exscale}
\usepackage{float}
\usepackage{bbm}
\usepackage[numbers,sort&compress]{natbib}
\newcommand{\N}{{\mathbb{N}}}
\newcommand{\Z}{{\mathbb{Z}}}

\newcommand{\CP}{{\mathbb{C}P}}
\newcommand{\1}{{\mathbbm{1}}}
\newcommand{\p}{\partial}

\newcommand{\ontopof}[2]{\genfrac{}{}{0pt}{}{#1}{#2}}

\makeatletter
\@addtoreset{equation}{section}
\makeatother

\title{Symmetry Analysis of Holes Localized on a Skyrmion in a Doped 
Antiferromagnet}

\author{N.\ D.\ Vlasii$^a$, C.\ P.\ Hofmann$^b$, F.-J.\ Jiang$^c$, and 
U.-J.\ Wiese$^d$
\\ \\
$^a$ Physics Department \\ 
Taras Shevchenko National University of Kyiv \\
64 Volodymyrska str., Kyiv 01601, Ukraine \\ \\
$^b$ Facultad de Ciencias, Universidad de Colima \\ 
Colima C.P.\ 28045, Mexico \\ \\
$^c$ Department of Physics, National Taiwan Normal University \\
88, Sec.\ 4, Ting-Chou Rd., Taipei 116, Taiwan \\ \\
$^d$ Albert Einstein Center for Fundamental Physics \\ 
Institute for Theoretical Physics, Bern University \\
Sidlerstrasse 5, CH-3012 Bern, Switzerland}

\begin{document} 

\maketitle

\begin{abstract} \normalsize

We use the low-energy effective field theory for holes coupled to the staggered
magnetization in order to investigate the localization of holes on a Skyrmion in
a square lattice antiferromagnet. When two holes get localized on the same 
Skyrmion, they form a bound state. The quantum numbers of the bound state are 
determined by the quantization of the collective modes of the Skyrmion. 
Remarkably, for p-wave states the quantum numbers are the same as those of a 
hole-pair bound by one-magnon exchange. Two holes localized on a Skyrmion with 
winding number $n = 1$ or 2 may have s- or d-wave symmetry as well. Possible 
relations with preformed Cooper pairs of high-temperature superconductors are 
discussed.

\end{abstract}
 
\maketitle
 
\newpage

\section{Introduction}

In the cuprates, high-temperature superconductivity is separated from 
antiferromagnetism by a pseudo-gap regime. It has been conjectured that the
relevant low-energy degrees of freedom in the pseudo-gap regime are responsible
for superconductivity as well. Reliably identifying those degrees of freedom by
theoretical investigations is a highly non-trivial task, because unbiased first
principles analytic or numerical calculations in microscopic systems such as 
the Hubbard or $t$-$J$ model are presently out of reach. In lightly doped 
antiferromagnets, on the other hand, the situation is more favorable. First of 
all, precise numerical simulations of undoped antiferromagnets 
\cite{Wie94,Bea96,San10,Jia11},(such as the Heisenberg model) are possible with 
the loop-cluster algorithm \cite{Eve93}, and individual doped holes can also be 
simulated reliably \cite{Bru00,Mis01}. Second, the low-energy dynamics of 
lightly doped 
antiferromagnets can be described with a systematic effective field theory for 
magnons and holes. The pure magnon effective field theory has been developed in
\cite{Cha89,Neu89,Fis89,Has90} and is completely analogous to chiral 
perturbation theory for the Goldstone pions in QCD \cite{Gas85}. In the past few
years, the systematic effective theory for magnons and doped holes has been 
constructed \cite{Kae05,Bru06} in complete analogy to baryon chiral perturbation
theory --- the effective theory for pions and nucleons 
\cite{Gas88,Jen91,Ber92,Bec99}. In contrast to previous attempts to construct 
effective theories for magnons and holes in a square lattice antiferromagnet
\cite{Shr88,Wen89,Sha90,Kue93,Kuc93}, the construction of \cite{Bru06} is
based on a systematic symmetry analysis and provides a complete set of all 
terms contributing to the effective action at leading and sub-leading order. As 
a result, the predictions of the effective theory are exact, order by order in 
a systematic derivative expansion. In particular, the low-energy physics of any 
lightly doped antiferromagnet is described quantitatively once some low-energy 
parameters (such as the spin stiffness or the spinwave velocity of the 
underlying microscopic system) have been fixed either by experiment or by 
numerical simulations. The effective theory has been used in systematic studies
of magnon-mediated two-hole bound states \cite{Bru05} and of spiral phases 
\cite{Bru07}. Earlier (but somewhat less systematic) studies had been presented 
in \cite{Kuc93,Sus04,Kot05}. Systematic effective field theories have also been 
constructed for antiferromagnets on a honeycomb lattice \cite{Kae11,Jia09}, as
well as for lightly electron-doped antiferromagnets \cite{Bru07a}.

Unfortunately, before one enters the high-temperature superconductor or even
just the pseudo-gap regime, both antiferromagetism and the systematic effective
theory that describes it break down. While one might expect that one can hence
not learn anything about high-temperature superconductivity or the pseudo-gap 
regime from the effective theory, the situation may not be entirely hopeless.
In particular, the effective theory still contains information about what 
objects may form when the theory is about to break down. In this way, we can 
identify new candidate low-energy degrees of freedom for which another 
effective theory with an extended validity range can be constructed. In this
paper, we do not yet attempt to construct an effective field theory for the
pseudo-gap regime. Instead, we concentrate on the identification of new 
low-energy objects that may form when antiferromagnetism is about to break 
down. 

When antiferromagnetism is weakened, the spin stiffness $\rho_s$ is reduced. In
particular, if antiferromagnetism is ultimately destroyed in a second order 
phase transition, $\rho_s$ vanishes at the transition. A small value of 
$\rho_s$ favors topological excitations in the staggered magnetization --- the 
order parameter for antiferromagnetism. In $(2+1)$ dimensions, the topological
excitations of the staggered magnetization vector are Skyrmions which
carry a topologically conserved winding number $n \in \Pi_2[S^2] = \Z$ in the
second homotopy group of the order parameter manifold $S^2$. The coset space
$S^2 = SU(2)_s/U(1)_s$ arises because in an antiferromagnet the $SU(2)_s$ spin
symmetry is spontaneously broken down to the subgroup $U(1)_s$. The possible 
role of Skyrmions as relevant excitations in quantum antiferromagnets has been 
discussed in several publications \cite{Hal88,Rea89,Shr90,Goo91,Goo93,Haa96,Mar01,Mot03,Sen04,Bae04,Wie05,Mor05,Fu10,Rai11,Bas11}. Haldane was first to realize 
that Skyrmions in an antiferromagnet are associated with a geometric phase 
\cite{Hal88}. When Skyrmions proliferate, antiferromagnetic order is destroyed. 
Read and Sachdev showed that on a square lattice the Skyrmion's geometric phase 
then implies a competing valence bond solid order with 4-fold degeneracy
\cite{Rea89}. The interplay of geometric phases and competing orders has been
discussed in detail in \cite{Fu10}. The suppression of Skyrmions has been 
related to unconventional deconfined quantum critical points \cite{Mot03,Sen04}.
It has also been argued that a hole localized near a dopant stabilizes a 
Skyrmion texture in the staggered magnetization 
\cite{Shr90,Goo91,Goo93,Haa96,Mar01}. The analogies between pions in QCD and 
magnons in ferro- and antiferromagnets have been investigated in detail in
\cite{Bae04,Wie05}. In particular, it was argued that Skyrmions endowed with 
fermion number 2 may act as preformed Cooper pairs of high-temperature
superconductivity. Experimental evidence for Skyrmions in the lightly doped 
insulating antiferromagnet $\mbox{La}_2 \mbox{Cu}_{1-x} \mbox{Li}_x \mbox{O}_4$ 
in an external magnetic field has been reported in \cite{Rai11}. Furthermore,
the possible role of Skyrmions for the superconductivity of $\mbox{Fe}$ based 
pnictides and chalogenides has been discussed in \cite{Bas11}. In this paper, 
for the first time, we investigate the localization of holes on a Skyrmion 
using the low-energy effective theory for lightly hole-doped antiferromagnets
on a square lattice. In particular, we carefully quantize the Skyrmion's 
collective modes, which allows us to unambiguously determine the quantum numbers
of single holes as well as hole pairs localized on a Skyrmion.

At the classical level, the mass of a Skyrmion is given by $4 \pi \rho_s$. When
$\rho_s$ becomes small, these excitations hence become energetically favorable.
Skyrmions are beyond reach of the systematic derivative expansion of the
low-energy effective theory for magnons and holes. Indeed, when Skyrmions
become relevant low-energy degrees of freedom, antiferromagnetism as well as 
the effective theory that describes it are about to break down. Still, the
effective theory correctly describes the way in which holes couple to a 
Skyrmion excitation in the staggered magnetization order parameter. In
particular, holes may get localized on a Skyrmion. When two holes get 
localized on the same Skyrmion, they form a bound state which may 
represent a relevant low-energy degree of freedom even when antiferromagnetism
gives way to the pseudo-gap phase. In particular, such bound states are a 
potential candidate for preformed pairs whose condensation may ultimately lead 
to high-temperature superconductivity. In order to decide whether this is a 
viable scenario, in this paper we investigate the symmetry properties of 
Skyrmion-hole bound states in great detail. We find that the p-wave states
of two holes localized on a Skyrmion with winding number $n = 1$ transform
exactly like the two-hole states weakly bound by one-magnon exchange. Two
holes localized on a Skyrmion may also have s- or d-wave symmetry. Which of 
these states is energetically most favorable depends on the details of the 
dynamics, and will remain a subject for future investigations.

The rest of the paper is organized as follows. In Section 2 the effective theory
for the staggered magnetization order parameter is introduced and Skyrmions are
discussed as classical solutions. The Hopf term is introduced and the collective
modes of a rotating Skyrmion are then quantized. In Section 3 doped holes are 
added to the effective theory. In Section 4 states of single holes as well as a 
pair of holes (residing in two different hole pockets) localized on a static or 
rotating Skyrmion are constructed and their symmetry properties are 
investigated. Possible relations to the mechanism responsible for 
high-temperature superconductivity are also discussed. Section 5 contains our 
conclusions. Finally, the case of two holes residing in the same hole pocket is
investigated in Appendix A.

\section{Skyrmions in the Effective Theory for the Staggered Magnetization}

In this section we discuss the collective mode quantization of Skyrmions in
the low-energy effective theory for antiferromagnetic magnons. 

\subsection{Effective Action and its Symmetries}

Magnons are the Goldstone bosons of a spontaneously broken spin symmetry 
$SU(2)_s$ with an unbroken subgroup $U(1)_s$. Consequently, magnons are 
described by a 3-component unit-vector field $\vec e(x) \in S^2$ in the coset 
space $S^2 = SU(2)_s/U(1)_s$. Here $x = (x_1,x_2,t)$ is a point in $(2+1)$-d 
Euclidean space-time and $\vec e(x)$ represents the direction of the local
staggered magnetization vector --- the order parameter for the spontaneously
broken spin symmetry. To leading order in a systematic derivative expansion, 
the Euclidean low-energy effective action for the magnons is given by
\begin{equation}
\label{effact}
S[\vec e] = \int d^2x \ dt \ \frac{\rho_s}{2}
\left(\p_i \vec e \cdot \p_i \vec e +
\frac{1}{c^2} \p_t \vec e \cdot \p_t \vec e\right).
\end{equation}
Here $\rho_s$ is the spin stiffness and $c$ is the spinwave velocity. The 
vacuum configuration of the effective theory is described by a constant 
staggered magnetization vector which can be chosen to point in the 
$3$-direction, i.e.\ $\vec e(x) = (0,0,1)$. Magnons are small fluctuations 
around the vacuum configuration. It should be noted that, in contrast to a
ferromagnet, antiferromagnetic magnons have a ``relativistic'' dispersion 
relation.

The most important symmetry of the action is the spontaneously broken spin
symmetry $SU(2)_s$. In the following, global transformations in the unbroken 
subgroup $U(1)_s$ will play an important role. Introducing
\begin{equation}
\vec e(x) = (\sin\theta(x) \cos\varphi(x),\sin\theta(x) \sin\varphi(x),
\cos\theta(x)),
\end{equation}
these transformations take the form
\begin{equation}
\label{subgroup}
^{I(\gamma)}\vec e(x) = (\sin\theta(x) \cos(\varphi(x) + \gamma),
\sin\theta(x) \sin(\varphi(x) + \gamma),\cos\theta(x)).
\end{equation}
It should be pointed out that the $SU(2)_s$ spin symmetry plays the role of an 
internal symmetry (analogous to chiral symmetry in particle physics). 
Consequently, its unbroken $U(1)_s$ subgroup (which is analogous to isospin in 
particle physics) should also be viewed as an internal symmetry. Because of the
analogy with isospin, we denote transformations in the unbroken subgroup 
$U(1)_s$ by $I(\gamma)$.

In addition to the $SU(2)_s$ spin symmetry, the effective action has other 
symmetries as well. First of all, due to the relativistic dispersion relation 
of antiferromagnetic magnons, the leading terms in the effective action have 
an emergent accidental Poin\-ca\-r\'e symmetry which is not present in the 
underlying Hubbard or $t$-$J$ model, and which will thus be explicitly broken 
by higher-order terms in the effective action containing a larger number of 
derivatives. The remaining symmetries are the discrete translations and 
rotations of the underlying quadratic lattice. Similar to the spin symmetry, 
the displacements $D_i$ by one lattice spacing in the $i$-direction are also 
spontaneously broken in an antiferromagnet. They act on the staggered 
magnetization field as
\begin{equation}
^{D_i}\vec e(x) = - \vec e(x).
\end{equation}
Since the shift symmetries $D_i$ are spontaneously broken in an 
antiferromagnet, it is convenient to also introduce modified shift symmetries
$D_i'$ which combine $D_i$ with an $SU(2)_s$ spin rotation $g = i \sigma_2$ 
such that
\begin{equation}
^{D_i'}\vec e(x) = (e_1(x),- e_2(x),e_3(x)).
\end{equation}
Spatial translations by an even number of lattice spacings, on the other hand, 
remain unbroken. Such translations $D(x_0)$ by a distance vector 
$x_0 = (x_{01},x_{02},0)$ act as
\begin{equation}
^{D(x_0)}\vec e(x) = \vec e(x - x_0).
\end{equation}
Similarly, parametrizing $x = (r \cos\chi,r \sin\chi,t)$, spatial rotations by 
an angle $\beta$ act as
\begin{equation}
^{O(\beta)}\vec e(x) = \vec e(O(\beta)x), \ 
O(\beta)x = (r \cos(\chi + \beta),r \sin(\chi + \beta),t),
\end{equation}
and a spatial reflection at the $x_1$-axis is represented by
\begin{equation}
^R\vec e(x) = \vec e(Rx), \ Rx = (x_1,-x_2,t) = (r \cos\chi,- r \sin\chi,t).
\end{equation}
Finally, time reversal, which changes the direction of a spin, acts as
\begin{equation}
^T\vec e(x) = - \vec e(Tx), \ Tx = (x_1,x_2,-t) = (r \cos\chi,r \sin\chi,-t).
\end{equation}
The effective action of eq.(\ref{effact}) is invariant under all these
symmetries.

\subsection{Classical Skyrmion Solutions}

In particle physics Skyrmions arise as topological excitations in the pion 
effective field theory for the strong interactions \cite{Sky61}, which takes 
the form of a $(3+1)$-d $SU(2)_L \times SU(2)_R = O(4)$ model. In order to 
distinguish them from their particle physics analogs, the topological 
excitations in the $(2+1)$-d $O(3)$ model are sometimes denoted as 
baby-Skyrmions. For simplicity, here we also refer to them just as Skyrmions. 
Skyrmions are topologically non-trivial classical solutions of the magnon 
effective theory with integer winding number
\begin{equation}
n[\vec e] = \frac{1}{8 \pi} \int d^2x \ \varepsilon_{ij} \vec e \cdot 
\left[\p_i \vec e \times \p_j \vec e\right] \in \Pi_2[S^2] = \Z,
\end{equation}
in the second homotopy group of the sphere $S^2$. Correspondingly, there is a
topological current
\begin{equation}
j_\mu(x) = \frac{1}{8 \pi} \varepsilon_{\mu\nu\rho} \vec e(x) \cdot 
\left[\p_\nu \vec e(x) \times \p_\rho \vec e(x)\right],
\end{equation}
which is conserved, i.e.\ $\p_\mu j_\mu(x) = 0$, irrespective of the classical
equations of motion. The winding number $n[\vec e] = \int d^2x \ j_t(x)$ is
just the integrated topological charge density. Under the various symmetries 
the topological charge density transforms as
\begin{eqnarray}
U(1)_s:&&^{I(\gamma)}j_t(x) = j_t(x), \nonumber \\
D_i:&&^{D_i}j_t(x) = - j_t(x), \nonumber \\
D_i':&&^{D_i'}j_t(x) = - j_t(x), \nonumber \\
O(\beta):&&^{O(\beta)}j_t(x) = j_t(O(\beta)x), \nonumber \\
R:&&^Rj_t(x) = - j_t(Rx), \nonumber \\
T:&&^Tj_t(x) = - j_t(Tx).
\end{eqnarray}
In particular, the winding number changes sign under the displacements $D_i$ 
and $D_i'$ as well as under the reflection $R$ and under the time reversal $T$.

Let us consider static classical solutions for which the energy
\begin{equation}
E[\vec e] = \int d^2x \ \frac{\rho_s}{2} \p_i \vec e \cdot \p_i \vec e
\end{equation}
is minimized. We can write
\begin{eqnarray}
0&\leq&\int d^2x \ \left(\p_i \vec e \pm 
\varepsilon_{ij} \p_j \vec e \times \vec e\right)^2 \nonumber \\
&=&\int d^2x \ \left(2 \p_i \vec e \cdot \p_i \vec e \pm 2
\varepsilon_{ij} \vec e \cdot \left(\p_i \vec e \times \p_j \vec e\right)\right)
= \frac{4}{\rho_s} E[\vec e] \pm 16 \pi n[\vec e],
\end{eqnarray}
which implies the Schwarz inequality
\begin{equation}
\label{schwarz}
E[\vec e] \geq 4 \pi \rho_s |n[\vec e]|.
\end{equation}
Skyrmions are minima of the energy in the topological sector with 
$n[\vec e] = 1$, while anti-Skyrmions have $n[\vec e] = - 1$. At the classical 
level both have a rest energy of ${\cal M} c^2 = 4 \pi \rho_s$. 
(Anti-)Skyrmions satisfy the previous inequality as an equality which is 
possible only if they satisfy the (anti-)self-duality equation
\begin{equation}
\p_i \vec e + \sigma \varepsilon_{ij} \p_j \vec e \times \vec e = 0.
\end{equation}
Here $\sigma = \pm 1$ distinguishes between Skyrmions and anti-Skyrmions. It is 
worth mentioning that static (anti-)Skyrmions are mathematically equivalent to 
(anti-)in\-stan\-tons of the 2-d $O(3)$ model \cite{Bel75}. Using polar 
coordinates $(x_1,x_2) = r (\cos\chi,\sin\chi)$, a particular (anti-)Skyrmion 
configuration is given by
\begin{equation}
\label{skyrmion}
\vec e_{\sigma,n,\rho}(r,\chi) = \left(\frac{2 r^n \rho^n}{r^{2n} + \rho^{2n}} 
\cos(n \chi),\frac{2 r^n \rho^n \sigma}{r^{2n} + \rho^{2n}} \sin(n \chi),
\frac{r^{2n} - \rho^{2n}}{r^{2n} + \rho^{2n}}\right).
\end{equation}
Depending on the sign of $\sigma$, this configuration describes a Skyrmion or
anti-Skyrmion of winding number $n[\vec e] = \sigma n$ (with $n \in \N_{>0}$) 
and size $\rho$ centered at the 
origin. It should be noted that there are many other multi-Skyrmion 
configurations with different Skyrmions located in different positions. Such 
configurations would be important in investigations of a Skyrmion gas or 
liquid. Here we concentrate on a Skyrmion centered at a single point, possibly 
with a larger winding number than just $n = 1$. The winding is chosen to arise 
from the angular $\chi$-dependence which influences the rotational symmetry of 
the Skyrmion and not from the radial $r$-dependence which only influences the 
finer details of the dynamics.

The Skyrmion configurations of eq.(\ref{skyrmion}) have a number of zero-modes.
In particular, their energy remains unchanged when they are shifted to an 
arbitrary position $x$, when they are spatially rotated by an arbitrary angle 
$\beta$, or when they are $U(1)_s$ spin-rotated by an arbitrary angle $\gamma$.
Interestingly, spatial rotations and $U(1)_s$ spin rotations act on a Skyrmion 
in a similar manner, i.e.
\begin{eqnarray}
^{O(\beta)}\vec e_{\sigma,n,\rho}(r,\chi)\!\!\!\!&=&\!\!\!\!
\left(\frac{2 r^n \rho^n}
{r^{2n} + \rho^{2n}} \cos(n (\chi + \beta)),
\frac{2 r^n \rho^n \sigma}{r^{2n} + \rho^{2n}} \sin(n (\chi + \beta)),
\frac{r^{2n} - \rho^{2n}}{r^{2n} + \rho^{2n}}\right), \nonumber \\
^{I(\sigma \gamma)}\vec e_{\sigma,n,\rho}(r,\chi)\!\!\!\!&=&\!\!\!\!
\left(\frac{2 r^n \rho^n}
{r^{2n} + \rho^{2n}} \cos(n \chi + \gamma),
\frac{2 r^n \rho^n \sigma}{r^{2n} + \rho^{2n}} \sin(n \chi + \gamma),
\frac{r^{2n} - \rho^{2n}}{r^{2n} + \rho^{2n}}\right), \nonumber \\ \,
\end{eqnarray}
such that
\begin{equation}
\label{rotations}
^{I(\sigma \gamma)}\vec e_{\sigma,n,\rho}(r,\chi) = \, 
^{O(\gamma/n)}\vec e_{\sigma,n,\rho}(r,\chi).
\end{equation}
Another zero-mode is related to dilations. Indeed, the energy of a Skyrmion 
also remains invariant under changes of the scale parameter $\rho$. A family of
Skyrmion configurations is obtained by spin-rotating the original Skyrmion of
eq.(\ref{skyrmion}) by an angle $\sigma \gamma$ and then shifting it by a 
distance-vector $x$ such that
\begin{equation}
\label{Skyrmionrot}
\vec e_{\sigma,n,\rho,x,\gamma}(r,\chi) = \,
^{D(x)}\left[^{I(\sigma \gamma)}\vec e_{\sigma,n,\rho}(r,\chi)\right].
\end{equation}
Under the various unbroken symmetry transformations the configuration of 
eq.(\ref{Skyrmionrot}) transforms as
\begin{alignat}{2}
U(1)_s:&\quad &^{I(\sigma \gamma_0)}\vec e_{\sigma,n,\rho,x,\gamma}(r,\chi) &= 
\vec e_{\sigma,n,\rho,x,\gamma + \gamma_0}(r,\chi), \nonumber \\
D'_i:&\quad &^{D_i'}\vec e_{\sigma,n,\rho,x,\gamma}(r,\chi) &=
\vec e_{- \sigma,n,\rho,x,\gamma}(r,\chi), \nonumber \\
D:&\quad &^{D(x_0)}\vec e_{\sigma,n,\rho,x,\gamma}(r,\chi) &=
\vec e_{\sigma,n,\rho,x + x_0,\gamma}(r,\chi), \nonumber \\
O(\beta):&\quad &^{O(\beta)}\vec e_{\sigma,n,\rho,x,\gamma}(r,\chi) &=
\vec e_{\sigma,n,\rho,O(\beta)x,\gamma + n \beta}(r,\chi), \nonumber \\
R:&\quad &^R\vec e_{\sigma,n,\rho,x,\gamma}(r,\chi) &= 
\vec e_{- \sigma,n,\rho,Rx,- \gamma}(r,\chi).
\end{alignat}

In particle physics Skyrmions play an interesting role in the effective theory 
for the strong interactions. In particular, Skyrmions arise as topological
excitations in the pion field \cite{Sky61}. While Skyrmions are outside the 
validity range of the systematic low-energy expansion of chiral perturbation 
theory, they have been used to model baryons phenomenologically \cite{Adk83}.
Remarkably, the $\Pi_3[S^3]$ topological winding number of the Skyrmions of 
the strong interactions has the same symmetry properties as the baryon number,
and is indeed identified with it. The identification of Skyrmions as baryons
can even be established within the framework of chiral perturbation theory, by
investigating the electromagnetic interactions of pions which are affected
by a Goldstone-Wilczek current \cite{Gol81,Bae01,Bae04,Wie05}. Since the 
underlying QCD theory has a conserved baryon number current, the conservation of
the topological Skyrme current is guaranteed beyond the semi-classical regime. 

It is natural to ask whether the winding number $n[\vec e] \in \Pi_2[S^2]$ of 
the Skyrmions in an antiferromagnet can also be identified with a conserved
quantity of an underlying microscopic system, such as the Hubbard model. In
particular, in analogy to particle physics, one might suspect that the winding
number can be identified with the fermion number of doped holes. However, this
is not the case because the winding number and the fermion number have 
different symmetry properties. In particular, the winding number changes sign
under a shift $D_i$ by one lattice spacing, while the fermion number does not.
Hence, unlike in particle physics, in an antiferromagnet the conservation of 
the topological current is not protected by the underlying microscopic dynamics
and may thus be limited to the semi-classical regime. Interestingly, when holes
get localized on a Skyrmion, they endow the Skyrmion with their conserved 
fermion number, which may stabilize the Skyrme beyond the semi-classical 
regime. 

The conservation of the topological current also plays a central role in the 
scenario of deconfined quantum criticality \cite{Sen04} in which dynamically
generated gauge fields and deconfined spinons are conjectured to appear at a 
new type of quantum phase transition outside the realm of the standard 
Ginsburg-Landau-Wilson paradigm. In fact, the suppression of Skyrmion number 
violating (so-called monopole) events has been argued to change the 
universality class of the phase transition in the $(2+1)$-d $O(3)$ model 
\cite{Mot03}. A better understanding of the role of Skyrmions would thus also
be useful for addressing the issue of deconfined quantum criticality.

\subsection{The Hopf Term}

The integer winding number $n[\vec e]$ is defined at any instant of time and is
conserved for topological reasons. Interestingly, there is another topological
invariant --- the Hopf number $H[\vec e]$ --- which characterizes the
topology of the order parameter field $\vec e(x)$ as a function of both space
and time. The integer-valued Hopf number $H[\vec e] \in \Pi_3[S^2] = \Z$ is an
element of the third homotopy group of the sphere $S^2$. In order to construct
the Hopf term, it is most convenient to introduce the $\CP(1)$ representation
\begin{equation}
P(x) = \frac{1}{2}\left(\1 + \vec e(x) \cdot \vec \sigma\right)
\end{equation}
of the staggered magnetization field. Here $\vec \sigma$ are the Pauli matrices
and, as a result, $P(x)$ is a Hermitean $2 \times 2$ projector matrix that 
obeys
\begin{equation}
P(x)^\dagger = P(x), \ P(x)^2 = P(x), \ \mbox{Tr} P(x) = 1.
\end{equation}
Under a spin rotation $g \in SU(2)_s$ the matrix $P(x)$ transforms as
\begin{equation}
P(x)' = g P(x) g^\dagger.
\end{equation}
The matrix $P(x)$ can be diagonalized by a unitary transformation 
$u(x) \in SU(2)$, i.e.
\begin{equation}
u(x) P(x) u(x)^\dagger = \frac{1}{2}(\1 + \sigma_3) = 
\left(\begin{array}{cc} 1 & 0 \\ 0 & 0 \end{array} \right), \qquad 
u_{11}(x) \geq 0.
\end{equation}
We demand that $u_{11}(x)$ is real and positive, which fixes a $U(1)_s$ 
gauge ambiguity and uniquely determines $u(x)$ as
\begin{eqnarray}
\label{defu}
u(x)&=&\frac{1}{\sqrt{2 (1 + e_3(x))}}
\left(\begin{array}{cc} 1 + e_3(x) & e_1(x) - i e_2(x) \\ 
- e_1(x) - i e_2(x) & 1 + e_3(x) \end{array}\right) \nonumber \\
&=&\left(\begin{array}{cc} \cos\left(\frac{1}{2} \theta(x)\right) & 
\sin\left(\frac{1}{2} \theta(x)\right) \exp(- i \varphi(x)) \\
- \sin\left(\frac{1}{2} \theta(x)\right) \exp(i \varphi(x)) & 
\cos\left(\frac{1}{2} \theta(x)\right) \end{array}\right) \nonumber \\
&=&\cos\left(\frac{1}{2} \theta(x)\right) + 
i \sin\left(\frac{1}{2} \theta(x)\right) 
\vec e_\varphi(x) \cdot \vec \sigma,
\end{eqnarray}
where the unit-vector $\vec e_\varphi(x)$ is given by
\begin{equation}
\vec e_\varphi(x) = \left(- \sin \varphi(x),\cos \varphi(x),0\right). 
\end{equation}
Under a global $SU(2)_s$ transformation $g$, the diagonalizing field $u(x)$
transforms as
\begin{equation}
\label{trafou}
u(x)' = h(x) u(x) g^\dagger, \qquad u_{11}(x)' \geq 0,
\end{equation}
which implicitly defines the nonlinear symmetry transformation 
\begin{equation}
h(x) = \exp(i \alpha(x) \sigma_3) = \left(\begin{array}{cc}
\exp(i \alpha(x)) & 0 \\ 0 & \exp(- i \alpha(x)) \end{array} \right) \in U(1)_s.
\end{equation}
In this way, the global transformations $g \in SU(2)_s$ of the spontaneously
broken non-Abelian spin symmetry ``disguise'' themselves as local 
transformations $h(x) \in U(1)_s$ of the unbroken subgroup. The global subgroup 
transformations $I(\gamma)$ introduced in eq.(\ref{subgroup}) simply lead to 
$\alpha(x) = - \gamma/2$.

The diagonalizing matrix $u(x)$ maps space-time onto the group manifold $S^3$ 
of $SU(2)_s$. When the $(2+1)$-d space-time is also compactified to $S^3$, one
can relate the Hopf number $H[\vec e] \in \Pi_3[S^2] = \Z$ to the topological 
winding number $W[u] \in \Pi_3[SU(2)_s] = \Pi_3[S^3] = \Z$, i.e.
\begin{equation}
\label{Hopf}
H[\vec e] = W[u] = \frac{1}{24 \pi^2} \int dt \ d^2x \ \varepsilon_{\mu\nu\rho} 
\mbox{Tr}\left[\left(u^\dagger \p_\mu u\right) \left(u^\dagger \p_\nu u\right)
\left(u^\dagger \p_\rho u\right)\right].
\end{equation}
It should be noted that the evaluation of eq.(\ref{Hopf}) requires some care.
In particular, due to the $U(1)_s$ gauge fixing $u_{11}(x) \geq 0$, $u(x)$
covers only an $S^2$ subspace of the $SU(2)_s$ group manifold $S^3$. This may
seem to imply that the winding number $W[u]$, which counts the number of times
the map $u(x)$ covers $S^3$, should vanish. However, this is not the case 
because $u(x)$ in eq.(\ref{defu}) is singular at the Skyrmion center where 
$e_3(x) = - 1$ (i.e.\ $\theta(x) = \pi$). The singularities which lie on a 
vortex line encircled by $\vec e_\varphi(x)$ contribute non-trivially to 
eq.(\ref{Hopf}). Alternatively, one may remove the singularities in $u(x)$ by 
undoing the $U(1)_s$ gauge fixing $u_{11}(x) \geq 0$, which implies that $u(x)$ 
extends to all of $S^3$. Then eq.(\ref{Hopf}) can be evaluated in a 
straightforward manner. The Hopf term is $SU(2)_s$-invariant because
\begin{equation}
W[u'] = W[h u g^\dagger] = W[h] + W[u] - W[g] = W[u].
\end{equation}
Here we have used $W[g] = 0$, which follows because $g$ is constant, and
$W[h] = 0$, which follows because the Abelian gauge transformations 
$h(x) \in U(1)_s$ are topologically trivial in three dimensions, i.e.\
$\Pi_3[U(1)_s] = \Pi_3[S^1] = \{0\}$. 

Under the various relevant symmetries the Hopf number transforms as
\begin{eqnarray}
U(1)_s:&&H[^{I(\gamma)}\vec e] = H[\vec e], \nonumber \\
D_i:&&H[^{D_i}\vec e] = H[\vec e], \nonumber \\
D_i':&&H[^{D_i'}\vec e] = H[\vec e], \nonumber \\
O(\beta):&&H[^{O(\beta)}\vec e] = H[\vec e], \nonumber \\
R:&&H[^R\vec e] = - H[\vec e], \nonumber \\
T:&&H[^T\vec e] = - H[\vec e].
\end{eqnarray}
The Hopf term gives rise to an additional factor $\exp(i \Theta H[\vec e])$ in
the Euclidean path integral with $\Theta$ being the anyon statistics angle. 
In systems with reflection or time-reversal symmetry, the value of $\Theta$ is 
hence limited to $0$ or $\pi$. As we will see, in these cases Skyrmions are 
quantized as bosons or fermions, respectively. In systems without reflection
and time-reversal symmetry, arbitrary values of $\Theta$ are allowed, and then
the Skyrmions may have any (neither integer nor half-integer) spin. By 
investigating field configurations in which two Skyrmions interchange their 
positions, one can also show that Skyrmions pick up a phase $\exp(i \Theta)$ 
and thus obey anyon statistics \cite{Wil83}. It should be noted that the Hopf 
term is expected to be absent in doped cuprates 
\cite{Wen88,Hal88,Dom88,Fra88,Rea89}, while it is known to be present, for 
example, in quantum Hall ferromagnets \cite{Tyc95,Tyc95a,Gir98,Bae04}. In
order to keep the discussion as general as possible, we will include the Hopf 
term, although in the cuprates one expects $\Theta = 0$.

\subsection{Collective Mode Quantization of the Skyrmion}

Let us now consider the collective mode quantization of the
Skyrmion. The main goal is to understand the quantum numbers of the quantized
Skyrmion, first of all in an undoped system. It should be pointed out that
Skyrmions in an undoped antiferromagnet are heavy objects whose pair-creation 
is suppressed at low temperatures. When antiferromagnetism is weakened by hole 
doping, the Skyrmion mass is reduced and, in addition, the 
holes may lower their mass by getting localized on a Skyrmion. This favors 
Skyrmion formation in doped antiferromagnets. A central goal of this paper is 
to understand the quantum numbers of the Skyrmion-hole bound states. In this 
subsection, we consider the collective mode quantization of a Skyrmion in the 
undoped system.

In order to perform the collective mode quantization, we consider the zero-mode
parameters $\rho(t)$, $x(t)$, and $\gamma(t)$ as functions of time. We now
evaluate the Euclidean action (including the Hopf term) for a time-dependent
Skyrmion and (after a somewhat lengthy but straightforward calculation) we 
obtain
\begin{equation}
\label{Skyrmeaction}
S[\vec e_{\sigma,n,\rho,x,\gamma}] + i \Theta H[\vec e_{\sigma,n,\rho,x,\gamma}] =
\int dt \ \left({\cal M} c^2 + \frac{{\cal M}}{2} \dot x^2 + 
\frac{{\cal D}(\rho)}{2} \dot \rho^2 + \frac{{\cal I}(\rho)}{2} \dot \gamma^2 +
i n \frac{\Theta}{2 \pi} \dot \gamma \right).
\end{equation}
Here the Skyrmion's rest energy is given by
\begin{equation}
{\cal M} c^2 = \rho_s  \int d^2x 
\frac{4 n^2 \rho^{2n}r^{2n-2}}{(r^{2n} + \rho^{2n})^2} = 4 \pi \rho_s n,
\end{equation}
which confirms that for self-dual solutions the Schwarz inequality of 
eq.(\ref{schwarz}) is obeyed as an equality. For $n > 1$ the Skyrmion's 
inertia against dilations takes the form
\begin{equation}
{\cal D}(\rho) = \frac{\rho_s}{c^2} \int d^2x 
\frac{4 n^2 r^{2n} \rho^{2n-2}}{(r^{2n} + \rho^{2n})^2} = 
\frac{\pi {\cal M}}{n\sin(\pi/n)}.  
\end{equation}
For $n = 1$ the integral is logarithmically infrared divergent. In a finite 
volume or in a
system with a finite density of Skyrmions, the infrared divergence may be 
regularized because the volume available to each Skyrmion becomes effectively 
finite. Indeed such effects are known to arise in the instanton gas of the 2-d 
$O(3)$ model \cite{Fat79,Ber79}. In this paper, we do not attempt to decide
whether the same happens in the $(2+1)$-d $O(3)$ model that is relevant here. 
We just regularize ${\cal D}(\rho)$ by an infra-red cut-off $R$ which may or 
may not be infinite such that for $n = 1$
\begin{equation}
{\cal D}(\rho) = 
\frac{4 \pi \rho_s}{c^2} \int_0^R dr \frac{2 r^3}{(r^2 + \rho^2)^2} = {\cal M} 
\left(\log\frac{R^2 + \rho^2}{\rho^2} - \frac{R^2}{R^2 + \rho^2}\right).
\end{equation}
Finally, the moment of inertia of the Skyrmion is given by
\begin{equation}
{\cal I}(\rho) = \frac{\rho_s}{c^2} \int d^2x 
\frac{4 r^{2n} \rho^{2n}}{(r^{2n} + \rho^{2n})^2} = 
\frac{{\cal D}(\rho) \rho^2}{n^2},
\end{equation}
which is affected by the same infrared divergence as ${\cal D}(\rho)$. Hence, 
although the Skyrmion has a finite mass (and can thus undergo translational 
motion), in the limit of an infinite infra-red cut-off $R$ it has an infinite 
moment of inertia ${\cal I}(\rho)$ and can thus not rotate.

From the Euclidean action of eq.(\ref{Skyrmeaction}) we read off the
real-time Lagrange function as
\begin{equation}
L = \frac{{\cal M}}{2} \dot x^2 +
\frac{{\cal D}(\rho)}{2} \left(\dot \rho^2 + 
\frac{\rho^2}{n^2} \dot \gamma^2\right) -
n \frac{\Theta}{2 \pi} \dot \gamma - {\cal M} c^2.
\end{equation}
In the next step, we consider the canonically conjugate momenta
\begin{equation}
p_i = \frac{\p L}{\p \dot x_i} = {\cal M} \dot x_i, \
p_\rho = \frac{\p L}{\p \dot \rho} = {\cal D}(\rho) \dot \rho, \
p_\gamma = \frac{\p L}{\p \dot \gamma} = 
\frac{{\cal D}(\rho) \rho^2 \dot \gamma}{n^2} - n \frac{\Theta}{2 \pi}.
\end{equation}
It should be noted that, at the classical level, the $\Theta$-term is 
suppressed because relative to $p_\gamma$ it is of order $\hbar$ (which we have
put to 1). The canonically conjugate momenta lead to the classical Hamilton 
function
\begin{equation}
{\cal H} = p_i \dot x_i + p_\rho \dot \rho + p_\gamma \dot \gamma - L =
{\cal M} c^2 + \frac{p_i^2}{2 {\cal M}} + \frac{1}{2 {\cal D}(\rho)}
\left[p_\rho^2 + \frac{n^2}{\rho^2}\left(p_\gamma + 
n \frac{\Theta}{2 \pi}\right)^2 \right].
\end{equation}
The momentum $p_i$, the spin $p_\gamma$, and the energy 
\begin{equation}
E =  \frac{1}{2 {\cal D}(\rho)}
\left[p_\rho^2 + \frac{n^2}{\rho^2}
\left(p_\gamma + n \frac{\Theta}{2 \pi}\right)^2 
\right] = \frac{{\cal D}(\rho)}{2} \dot \rho^2 + 
\frac{n^2}{2 {\cal D}(\rho) \rho^2}
\left(p_\gamma + n \frac{\Theta}{2 \pi}\right)^2
\end{equation}
of the coupled rotational and dilational motion are conserved quantities. The
last equality determines the size $\rho(t)$ of the Skyrmion as a function of 
time
\begin{equation}
t = \int_{\rho(0)}^{\rho(t)} d\rho \left[\frac{2 E}{{\cal D}(\rho)} -
\frac{n^2}{{\cal D}(\rho)^2 \rho^2}
\left(p_\gamma + n \frac{\Theta}{2 \pi}\right)^2 \right]^{-1/2}.
\end{equation}
When the Skyrmion is rotating (i.e.\ when 
$p_\gamma + n \frac{\Theta}{2 \pi} \neq 0$),
centrifugal forces lead to an unlimited increase of $\rho(t)$.

Upon canonical quantization the momentum $p_i$ and the spin $p_\gamma$ turn into
the operators
\begin{equation}
p_i = - i \p_{x_i}, \ p_\gamma = - i \p_\gamma,
\end{equation}
while the classical Hamilton function ${\cal H}$ turns into the quantum 
mechanical Hamiltonian
\begin{equation}
H = {\cal M} c^2 - \frac{1}{2 {\cal M}} \p_{x_i}^2 - 
\frac{1}{\sqrt{2 {\cal D}(\rho)}}
\left(\p_\rho^2 + \frac{1}{\rho} \p_\rho \right)
\frac{1}{\sqrt{2 {\cal D}(\rho)}} - \frac{n^2}{2 {\cal D}(\rho) \rho^2}
\left(\p_\gamma + i n \frac{\Theta}{2 \pi}\right)^2.
\end{equation}
The collective mode wave function of a Skyrmion or anti-Skyrmion with winding 
number $\sigma n$, momentum $p_i$, and spin $p_\gamma = \sigma m \in \Z$ takes 
the form
\begin{equation}
\Psi_{p,\sigma,n,m}(x,\rho,\gamma) = \exp(i p_i x_i) \exp(i \sigma m \gamma) 
\psi(\rho).
\end{equation}
The dilational part of the wave function solves the Schr\"odinger equation
\begin{equation}
\left[- \frac{1}{\sqrt{2 {\cal D}(\rho)}}
\left(\p_\rho^2 + \frac{1}{\rho} \p_\rho\right)\frac{1}{\sqrt{2 {\cal D}(\rho)}}
+ \frac{n^2}{2 {\cal D}(\rho) \rho^2}
\left(m + n \sigma \frac{\Theta}{2 \pi}\right)^2\right] \psi(\rho) = 
E \psi(\rho),
\end{equation}
which may again lead to an instability of a rotating Skyrmion against 
unlimited increase of its size $\rho$. As we will see later, localized holes 
prevent the increase of $\rho$ and thus stabilize the Skyrmion.

In the presence of the Hopf term the spin operator of the Skyrmion (which is
analogous to isospin in particle physics) is given by
\begin{equation}
I = \sigma \left(p_\gamma + n \frac{\Theta}{2 \pi}\right) = 
\sigma \left(- i \p_\gamma + n \frac{\Theta}{2 \pi}\right).
\end{equation}
The state $\Psi_{p,\sigma,n,m}(x,\rho,\gamma)$ hence has the ``isospin''
\begin{equation}
I \Psi_{p,\sigma,n,m}(x,\rho,\gamma) = 
\left(m + \sigma n \frac{\Theta}{2 \pi}\right) \Psi_{p,\sigma,n,m}(x,\rho,\gamma).
\end{equation}
In particular, for $\Theta = 0$ the ``isospin'' is an integer, while for
$\Theta = \pi$ it is a half-integer for odd $n$.

Let us also investigate the quantum numbers of the Skyrmion with respect to
spatial rotations. As a consequence of eq.(\ref{rotations}), the angular 
momentum $J$ is given by 
\begin{equation}
J = \sigma n I = n \left(p_\gamma + n \frac{\Theta}{2 \pi}\right) =
n \left(- i \p_\gamma + n \frac{\Theta}{2 \pi}\right),
\end{equation}
such that 
\begin{equation}
J \Psi_{p,\sigma,n,m}(x,\rho,\gamma) = 
n \left(\sigma m + n \frac{\Theta}{2 \pi}\right)
\Psi_{p,\sigma,n,m}(x,\rho,\gamma).
\end{equation}
Hence, for $\Theta = 0$ the Skyrmion has integer angular momentum and thus is a
boson, while for $\Theta = \pi$ the angular momentum is a half-integer and the 
Skyrmion is a fermion. Interestingly, in $(2+1)$ dimensions it is possible to
have particles of any (neither integer nor half-integer) angular momentum --- 
the anyons which arise for $\Theta \neq 0$ or $\pi$.

By construction, the Skyrmion state is also an eigenstate of the momentum 
operator with eigenvalue $p_i$. Under the modified shift symmetries $D_i'$ and
under the reflection $R$ the Skyrmion state transforms as
\begin{eqnarray}
&&U_{D_i'} \Psi_{p,\sigma,n,m}(x,\rho,\gamma) = \Psi_{p,-\sigma,n,m}(x,\rho,\gamma), 
\nonumber \\
&&U_R \Psi_{p,\sigma,n,m}(x,\rho,\gamma) = \Psi_{Rp,-\sigma,n,m}(x,\rho,\gamma),
\end{eqnarray}
where $R p = (p_1,- p_2)$ is the spatially reflected momentum. Here $U_{D_i'}$ 
and $U_R$ are unitary transformations representing the corresponding discrete 
symmetries in the Hilbert space of the collective modes of the Skyrmion. It 
should be noted that shifted or reflected Skyrmions (which have $\sigma = 1$) 
are actually anti-Skyrmions (with $\sigma = - 1$).

\section{Effective Action for Doped Holes}

In order to make the paper self-contained, in this section we review the main
features of the
effective field theory constructed in \cite{Bru06} which couples doped holes to 
the staggered magnetization order parameter.

\subsection{Nonlinear Realization of the $SU(2)_s$ Symmetry}

In order to couple holes to the staggered magnetization order parameter, a 
nonlinear realization of the spontaneously broken $SU(2)_s$ symmetry has been 
constructed in \cite{Kae05}. The global $SU(2)_s$ symmetry then manifests 
itself as a local $U(1)_s$ symmetry in the unbroken subgroup. This is analogous
to baryon chiral perturbation theory in which the spontaneously broken 
$SU(2)_L \otimes SU(2)_R$ chiral symmetry of QCD is implemented on the nucleon 
fields as a local $SU(2)_{L=R}$ transformation in the unbroken isospin subgroup.

The definition of the nonlinear realization of the $SU(2)_s$ symmetry is based
on the diagonalizing matrix $u(x)$ defined in eq.(\ref{defu}), which transforms
as
\begin{equation}
^{D_i'}u(x) = u(x)^*,
\end{equation} 
under the modified displacement symmetry $D_i'$. Introducing the traceless 
anti-Her\-mi\-te\-an field
\begin{equation}
v_\mu(x) = u(x) \p_\mu u(x)^\dagger,
\end{equation}
one obtains the following transformation rules
\begin{alignat}{3}
SU(2)_s:&\quad &v_\mu(x)' &= h(x) [v_\mu(x) + \p_\mu] h(x)^\dagger,
  \hspace{-5em} \nonumber \\
D'_i:&\quad &^{D'_i}v_\mu(x) &= v_\mu(x)^*, \nonumber \\
O:&\quad &^Ov_i(x) &= \varepsilon_{ij} v_j(Ox), \quad
  &^Ov_t(x) &= v_t(Ox), \nonumber \\
R:&\quad &^Rv_1(x) &= v_1(Rx), \quad &^Rv_2(x) &= - v_2(Rx),
  \quad ^Rv_t(x) = v_t(Rx).
\end{alignat}
Writing
\begin{equation}
v_\mu(x) = i v_\mu^a(x) \sigma_a, \qquad
v_\mu^\pm(x) = v_\mu^1(x) \mp i v_\mu^2(x),
\end{equation}
the field $v_\mu(x)$ decomposes into an Abelian ``gauge'' field $v_\mu^3(x)$ 
and two ``charged'' vector fields $v_\mu^\pm(x)$.

Using eq.(\ref{Skyrmionrot}), for a Skyrmion 
$\vec e_{\sigma,n,\rho,0,\gamma}(r,\chi)$ centered at $x = 0$ one obtains
\begin{eqnarray}
\label{vSkyrmion}
v^3_1(r,\chi)&=&- \frac{\sigma n \rho^{2n}}{r (r^{2n} + \rho^{2n})} \sin\chi, 
\nonumber \\
v^3_2(r,\chi)&=&\frac{\sigma n \rho^{2n}}{r (r^{2n} + \rho^{2n})} \cos\chi, 
\nonumber \\ 
v^3_t(r,\chi)&=&\frac{\sigma \rho^{2n}}{r^{2n} + \rho^{2n}} \dot \gamma, 
\nonumber \\
v^\pm_1(r,\chi)&=&\mp i \frac{n r^{n-1} \rho^n}{r^{2n} + \rho^{2n}} 
\exp(\mp i \sigma \left[(n+1) \chi + \gamma\right]), \nonumber \\
v^\pm_2(r,\chi)&=&\frac{\sigma n r^{n-1} \rho^n}{r^{2n} + \rho^{2n}} 
\exp(\mp i \sigma \left[(n+1) \chi + \gamma\right]), \nonumber \\
v^\pm_t(r,\chi)&=&\frac{\sigma r^n \rho^n}{r^{2n} + \rho^{2n}} 
\exp(\mp i \sigma (n \chi + \gamma)) \dot \gamma.
\end{eqnarray}
In principle, when holes get localized on a Skyrmion, they affect the radial 
profile of the Skyrmion. Here we neglect this effect and concentrate on symmetry
considerations which are independent of such details of the dynamics.

\subsection{Hole Fields and their Transformation Properties}

As discussed in detail in \cite{Bru06} the holes are described by 
Grassman-valued fields $\psi^f_\pm(x)$. Here $f \in \{\alpha,\beta\}$ is a
flavor index which specifies the momentum space pocket in which the hole
resides, and the subscript $\pm$ denotes the spin of the hole relative to the
direction of the local staggered magnetization. Under the various relevant
symmetries of the underlying antiferromagnet on a square lattice, the hole 
fields transform as
\begin{alignat}{2}
\label{symcomp}
SU(2)_s:&\quad &\psi^f_\pm(x)' &= \exp(\pm i \alpha(x)) \psi^f_\pm(x),
\nonumber \\
U(1)_Q:&\quad &^Q\psi^f_\pm(x) &= \exp(i \omega) \psi^f_\pm(x),
\nonumber \\
D'_i:&\quad &^{D'_i}\psi^f_\pm(x) &= \pm \exp(i k^f_i a) \psi^f_\mp(x),
\nonumber \\
O:&\quad &^O\psi^\alpha_\pm(x) &= \mp \psi^\beta_\pm(Ox), \quad
^O\psi^\beta_\pm(x) = \psi^\alpha_\pm(Ox), \nonumber \\
R:&\quad &^R\psi^\alpha_\pm(x) &= \psi^\beta_\pm(Rx), \quad\;\;\;
^R\psi^\beta_\pm(x) = \psi^\alpha_\pm(Rx).
\end{alignat}
The $U(1)_Q$ symmetry is just fermion number, while 
$k^\alpha = (\frac{\pi}{2a},\frac{\pi}{2a})$ and 
$k^\beta = (\frac{\pi}{2a},- \frac{\pi}{2a})$ (with $a$ being the lattice 
spacing) point to the centers of the two hole pockets illustrated in figure 1. 
It is interesting that in the effective theory momentum indices of the 
underlying microscopic dynamics turn into internal flavor quantum numbers.
\begin{figure}[t]
\begin{center}
\vspace{-0.4cm}
\epsfig{file=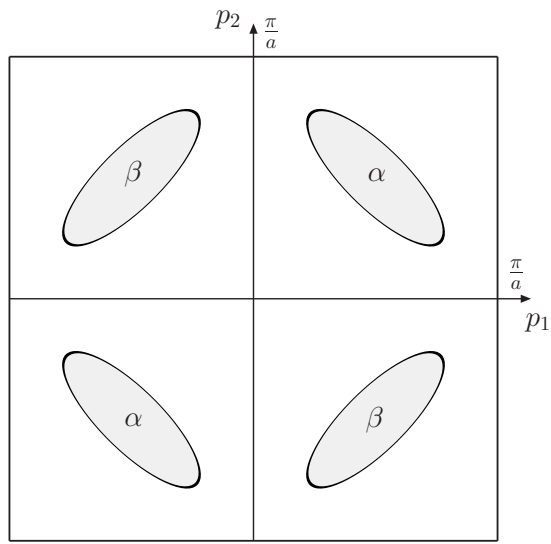,width=8cm}
\end{center}
\caption{\it Elliptically shaped hole pockets centered at 
$(\pm \frac{\pi}{2a},\pm \frac{\pi}{2a})$. Two half-pockets combine to form the
pockets for the flavors $f = \alpha,\beta$.}
\end{figure}

\subsection{Effective Action for  Holes coupled to the Staggered Magnetization}

Based on the above symmetry properties, the leading and sub-leading terms
of the effective action for an antiferromagnet on a square lattice have been 
constructed systematically in \cite{Bru06}. 
Here we restrict ourselves to the leading terms. We also make the simplifying
(but somewhat unrealistic) assumption that the momentum-space hole pockets have
a circular shape, which enables us to perform large parts of the following
calculations analytically. It would be straightforward to take into account the
more realistic elliptic shape of the hole pockets, but this would require some
numerical work. Here we concentrate foremost on the symmetry properties of 
holes localized on a Skyrmion on which the simplifying assumption of spherical 
hole pockets has no effect. The total action of the coupled system including
doped holes then takes the form
\begin{eqnarray}
\label{holeaction}
&&S[\psi^{f\dagger}_\pm,\psi^f_\pm,\vec e] = \int d^2x \ dt \ 
\Big\{\frac{\rho_s}{2} \left(\p_i \vec e \cdot \p_i \vec e +
\frac{1}{c^2} \p_t \vec e \cdot \p_t \vec e\right) \nonumber \\
&&+ \sum_{\ontopof{f=\alpha,\beta}{\, s = +,-}} 
\Big[M \psi^{f\dagger}_s \psi^f_s + \psi^{f\dagger}_s D_t \psi^f_s + 
\frac{1}{2 M'} D_i \psi^{f\dagger}_s D_i \psi^f_s +
\Lambda \big(\psi^{f\dagger}_s v^s_1 \psi^f_{-s} 
+ \sigma_f \psi^{f\dagger}_s v^s_2 \psi^f_{-s} \big)\Big]\Big\}. \nonumber \\
\end{eqnarray}
Here $M$ and $M'$ are the rest energy and the kinetic mass of a hole, and 
$\Lambda$ is the hole-one-magnon coupling constant. The sign $\sigma_f$ is $+$ 
for $f = \alpha$ and $-$ for $f = \beta$. The covariant derivatives are given 
by
\begin{equation}
D_\mu \psi^f_\pm(x) = \left[\p_\mu \pm i v_\mu^3(x)\right] \psi^f_\pm(x).
\end{equation}
Remarkably, the Shraiman-Siggia term in the action, which is proportional to 
$\Lambda$,
contains just a single (uncontracted) spatial derivative. Due to the nontrivial
rotation properties of flavor, this term is still 90 degrees rotation 
invariant. Due to the small number of derivatives it contains, this term 
dominates the low-energy dynamics. In particular, it alone is responsible for 
one-magnon exchange between hole pairs \cite{Bru05,Bru06} as well as for 
potential spiral phases in the staggered magnetization order parameter
\cite{Bru07}. It is interesting to note that a similar term is absent in 
lightly electron-doped antiferromagnets \cite{Bru07a}, such that spiral phases 
do not arise in these systems.

\section{Hole Localization on a Skyrmion}

In this section we apply the effective theory of the previous section to the
localization of holes on a Skyrmion. First, we consider the localization of a
single hole first on a static and then on a rotating Skyrmion. Then the 
localization of two holes on the same Skyrmion is considered, and the symmetry
properties of the resulting two-hole bound states are analyzed.

\subsection{Single Hole Localized on a Static Skyrmion}

As we have seen, the moment of inertia ${\cal I}(\rho)$ of a Skyrmion with
$n = 1$ is logarithmically divergent in the infra-red. Unless the divergence is
regularized due to a finite spatial volume or the presence of other Skyrmions,
the Skyrmion then cannot rotate. In the interest of analytic solubility, and 
because we want to focus on symmetry aspects, we will no longer consider the 
translational and dilational motion of the Skyrmion. Instead, we fix the 
Skyrmion center at the origin $x = 0$ and we fix the Skyrmion size to a 
constant $\rho$. As we will see later, in the presence of holes, the energy of 
the Skyrmion-hole bound states is minimized for a particular value of $\rho$.

The wave function of a single hole localized on a Skyrmion takes the form
\begin{equation}
\Psi^f_{\sigma,n}(r,\chi) = \left(\begin{array}{c}
\Psi^f_{\sigma,n,+}(r,\chi) \\ \Psi^f_{\sigma,n,-}(r,\chi)
\end{array} \right),
\end{equation}
Omitting the constant rest energy $M$ of the holes, which just amounts to a 
constant energy shift, the corresponding Hamiltonian resulting from the action 
of eq.(\ref{holeaction}) is given by
\begin{eqnarray}
\label{hamiltonian}
H^f&=&\left(\begin{array}{cc} H^f_{++} & H^f_{+-} \\ 
H^f_{-+} & H^f_{--} \end{array} \right), \nonumber \\
H^f_{++}&=&- \frac{1}{2 M'} \left[\p_i + i v^3_i(x)\right]^2 =
- \frac{1}{2 M'} \left[\p_r^2 + \frac{1}{r} \p_r +
\frac{1}{r^2} \left(\p_\chi + i \frac{\sigma n \rho^{2n}}{r^{2n} + \rho^{2n}}
\right)^2 \right], \nonumber \\
H^f_{+-}&=&\Lambda (v_1^+(x) + \sigma_f v_2^+(x)) \nonumber \\
&=&\sqrt{2} \Lambda \sigma \sigma_f \frac{n r^{n-1} \rho^n}{r^{2n} + \rho^{2n}}
\exp\left(- i \sigma\left[(n + 1) \chi + \gamma + 
\sigma_f \frac{\pi}{4}\right]\right),
\nonumber \\
H^f_{-+}&=&\Lambda (v_1^-(x) + \sigma_f v_2^-(x)) \nonumber \\
&=&\sqrt{2} \Lambda \sigma \sigma_f \frac{n r^{n-1} \rho^n}{r^{2n} + \rho^{2n}}
\exp\left(i \sigma\left[(n + 1) \chi + \gamma + 
\sigma_f \frac{\pi}{4}\right]\right),
\nonumber \\
H^f_{--}&=&- \frac{1}{2 M'} \left[\p_i - i v^3_i(x)\right]^2 =
- \frac{1}{2 M'} \left[\p_r^2 + \frac{1}{r} \p_r +
\frac{1}{r^2} \left(\p_\chi - i \frac{\sigma n \rho^{2n}}{r^{2n} + \rho^{2n}}
\right)^2 \right]. \qquad
\end{eqnarray}
Using the explicit form of $v_i^3(x)$ and $v_i^\pm(x)$ for the Skyrmion of
eq.(\ref{vSkyrmion}) and making the ansatz
\begin{equation}
\label{ansatz1}
\Psi^f_{\sigma,m_+,m_-}(r,\chi) = \left(\begin{array}{c} \psi_{m_+,m_-,+}(r)
\exp\left(i \sigma [m_+ \chi - \frac{\gamma}{2} - 
\sigma_f \frac{\pi}{8}]\right) \\ \sigma \sigma_f \psi_{m_+,m_-,-}(r) 
\exp\left(i \sigma [m_- \chi + \frac{\gamma}{2} 
+ \sigma_f \frac{\pi}{8}]\right) \end{array} \right),
\end{equation}
with $m_- - m_+ = n + 1$, after some algebra one obtains the radial 
Schr\"odinger equation
\begin{equation}
H_r \psi_{m_+,m_-}(r) = 
\left(\begin{array}{cc} H_{r++} & H_{r+-} \\ H_{r-+} & H_{r--} \end{array} \right)
\left(\begin{array}{c} \psi_{m_+,m_-,+}(r) \\ \psi_{m_+,m_-,-}(r) \end{array} \right)
= E_{m_+,m_-} \psi_{m_+,m_-}(r),
\end{equation}
with
\begin{eqnarray}
&&H_{r++} = - \frac{1}{2 M'} \left[\p_r^2 + \frac{1}{r} \p_r -
\frac{1}{r^2} \left(m_+ + \frac{n \rho^{2n}}{r^{2n} + \rho^{2n}}\right)^2\right],
\nonumber \\
&&H_{r+-} = H_{r-+} = 
\sqrt{2} \Lambda \frac{n r^{n-1} \rho^n}{r^{2n} + \rho^{2n}}, \nonumber \\
&&H_{r--} =  - \frac{1}{2 M'} \left[\p_r^2 + \frac{1}{r} \p_r - 
\frac{1}{r^2} \left(m_- - \frac{n \rho^{2n}}{r^{2n} + \rho^{2n}}\right)^2\right].
\end{eqnarray}
It should be noted that the resulting radial Schr\"odinger equation is the same
for Skyrmions and anti-Skyrmions as well as for both flavors $f = \alpha,\beta$.
Interestingly, for odd $n$ and $m_- = - m_+ = (n+1)/2$, the two equations 
decouple. The equation that leads to a localized hole takes the form
\begin{equation}
\label{decoupled_equation}
\left[- \frac{1}{2 M'}\left(\p_r^2 + \frac{1}{r} \p_r - \frac{1}{r^2} 
\left(\frac{n + 1}{2} - \frac{n \rho^{2n}}{r^{2n} + \rho^{2n}}\right)^2 \right)
- \frac{\sqrt{2} \Lambda n r^{n-1} \rho^n}{r^{2n} + \rho^{2n}} \right] \psi(r)
= E \psi(r),
\end{equation}
where $\psi(r)$ is the linear combination
\begin{equation}
\label{decoupled_wave}
\psi(r) = \frac{1}{\sqrt{2}}\left(\psi_{m_+,m_-,+}(r) - \psi_{m_+,m_-,-}(r)\right).
\end{equation}
For even winding number $n$, on the other hand, the two equations do not 
decouple. In the following we will be most interested in Skyrmions (or 
anti-Skyrmions) with winding number $n = 1$.

In this paper, we concentrate on the symmetry properties of holes localized on
a Skyrmion, not paying much attention to finer details of the dynamics. Hence, 
here we do not solve the radial equation, which would be straightforward using
numerical methods. Still, we want to obtain at least a rough estimate for the 
ground state energy of a hole localized on a Skyrmion. For $n = 1$ the radial 
Schr\"odinger equation takes the form
\begin{equation}
\left[- \frac{1}{2 M'} \left(\p_r^2 + \frac{1}{r} \p_r\right) + V(r)\right] 
\psi(r) = E \psi(r),
\end{equation}
with the potential given by
\begin{equation}
V(r) = \frac{1}{2 M'} \frac{r^2}{(r^2 + \rho^2)^2} - 
\sqrt{2} \Lambda \frac{\rho}{r^2 + \rho^2}.
\end{equation}
At short distances, the potential can be approximated by a harmonic oscillator
\begin{equation}
V_{\text{approx}}(r) = - \frac{\sqrt{2} \Lambda}{\rho} + \frac{M'}{2} 
\left(\frac{1}{{M'}^2 \rho^4} + \frac{2 \sqrt{2} \Lambda}{M' \rho^3}\right) r^2 
+ {\cal O}(r^4),
\end{equation}
and hence, in a rather crude harmonic approximation, the ground state energy 
takes the form
\begin{equation}
E_0 = - \frac{\sqrt{2} \Lambda}{\rho} + 
\sqrt{\frac{1}{{M'}^2 \rho^4} + \frac{2 \sqrt{2} \Lambda}{M' \rho^3}} =
M' \Lambda^2 x \left(\sqrt{x^2 + 2 \sqrt{2} x} - \sqrt{2}\right), \quad 
x = \frac{1}{M' \Lambda \rho}.
\end{equation}
Minimizing the energy as a function of $x$ yields 
$x^3 + 3 \sqrt{2} x^2 + 4 x = \sqrt{2}$, which is solved by
\begin{eqnarray}
\label{rho}
&&x = \sqrt{\frac{2}{3}} \left[
\left(\frac{3 \sqrt{3}}{4} + \frac{\sqrt{11}}{4}\right)^{1/3} +
\left(\frac{3 \sqrt{3}}{4} + \frac{\sqrt{11}}{4}\right)^{-1/3}\right] - \sqrt{2}
\approx 0.271 \ \Rightarrow \nonumber \\
&&\rho \approx \frac{1}{0.271 M' \Lambda}.
\end{eqnarray}
This shows that the presence of the hole explicitly breaks the scale invariance
that led to the dilational instability of the pure Skyrmion. The resulting 
bound state with the strongest binding energy has
\begin{equation}
\label{bindingE}
E_0 = M' \Lambda^2 x \left(\sqrt{x^2 + 2 \sqrt{2} x} - \sqrt{2}\right) \approx
- 0.135 M' \Lambda^2.
\end{equation}
The potential $V(r)$ is shown in figure 2 together with its harmonic 
approximation and the corresponding ground state energy $E_0$. The figure 
implies that the true ground state energy is smaller than the harmonic
approximation suggests.
\begin{figure}[t]
\begin{center}
\epsfig{file=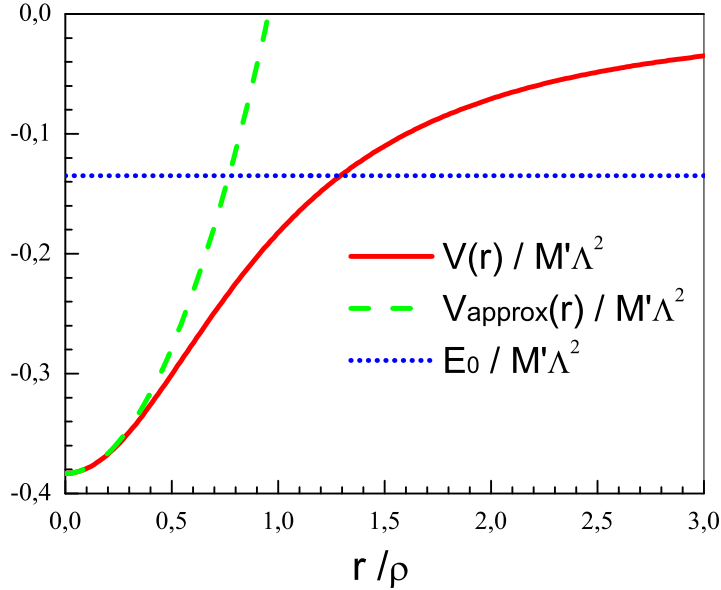,width=12cm}
\vspace{-3.5cm}
\end{center}
\caption{\it The potential $V(r)$ (solid curve) together with its harmonic 
approximation (dashed curve) and the corresponding ground state energy
(dotted line).}
\end{figure}

\subsection{Single Hole Localized on a Rotating Skyrmion}

In this subsection we consider a single hole localized on a rotating Skyrmion. 
When the moment of inertia ${\cal I}(\rho)$ diverges (as it is the case for 
$n = 1$ and $R = \infty$) the fixed orientation $\gamma$ of the Skyrmion 
explicitly breaks the $U(1)_s$ symmetry and the analysis of Section 4.1 
applies. Here we assume that ${\cal I}(\rho) = {\cal D}(\rho)\rho^2/n^2$ is 
finite. This is actually the case when the feedback of the localized hole on 
the radial structure of the Skyrmion is taken into account. When 
${\cal I}(\rho)$ is finite, the Skyrmion can rotate and thus $\gamma$ becomes a 
dynamical variable. The $\gamma$-dependent terms in the Lagrange function for 
the rotational motion are given by
\begin{equation}
L = \frac{{\cal D}(\rho) \rho^2}{2 n^2}\dot{\gamma}^2 -
n \frac{\Theta}{2 \pi} \dot \gamma +
\int d^{2}x \sum_{\ontopof{f=\alpha,\beta}{\, s = +,-}} 
s \psi^{f \dagger}_s v^3_t \psi^f_s. 
\end{equation}
Using eq.(\ref{vSkyrmion}), the momentum canonically conjugate to $\gamma$ thus
takes the form
\begin{equation}
p_\gamma = \frac{{\cal D}(\rho) \rho^2 \dot \gamma}{n^2} - 
n \frac{\Theta}{2 \pi} + \int d^{2}x \ \sigma 
\frac{\rho^{2n}}{r^{2n}+\rho^{2n}}
\sum_{\ontopof{f=\alpha,\beta}{\, s = +,-}} s \psi^{f \dagger}_s \psi^f_s,
\end{equation}
which leads to the corresponding Hamiltonian
\begin{equation}
H^\gamma = \frac{1}{2 {\cal I}(\rho)} 
\left(- i \partial_\gamma - A_\gamma\right)^2,
\end{equation} 
with the Berry gauge field
\begin{equation}
A_\gamma = \int d^2x \sum_{\ontopof{f=\alpha,\beta}{s = +,-}} 
\Psi^{f \dagger}_s
\frac{\sigma \rho^{2n}}{r^{2n}+\rho^{2n}} s \Psi^f_s - n \frac{\Theta}{2 \pi}.
\end{equation}
Combining the results, one sees that while the off-diagonal elements
of the Hamiltonian (\ref{hamiltonian}) remain the same, the diagonal elements 
receive additional contributions such that now
\begin{eqnarray}
H^f_{++}&=&- \frac{1}{2 M'} \left[\p_i + i v^3_i(x)\right]^2 - 
\frac{n^2}{2 {\cal D}(\rho) \rho^2} 
\left(\p_\gamma + i n \frac{\Theta}{2 \pi} -
i \sigma \frac{\rho^{2n}}{r^{2n} + \rho^{2n}}\right)^2 \nonumber \\
&=&- \frac{1}{2 M'} \left[\p_r^2 + \frac{1}{r} \p_r +
\frac{1}{r^2} \left(\p_\chi + i \sigma \frac{n \rho^{2n}}{r^{2n} + \rho^{2n}}
\right)^2 \right] \nonumber \\
&-&\frac{n^2}{2 {\cal D}(\rho) \rho^2} \left(\p_\gamma + 
i n \frac{\Theta}{2 \pi} - i \sigma \frac{\rho^{2n}}{r^{2n} + \rho^{2n}}
\right)^2, \nonumber \\
H^f_{+-}&=&\Lambda (v_1^+(x) + \sigma_f v_2^+(x)) \nonumber \\
&=&\sqrt{2} \Lambda \sigma \sigma_f \frac{n r^{n-1} \rho^n}{r^{2n} + \rho^{2n}}
\exp\left(- i \sigma\left[(n + 1) \chi + \gamma + 
\sigma_f \frac{\pi}{4}\right]\right),
\nonumber \\
H^f_{-+}&=&\Lambda (v_1^-(x) + \sigma_f v_2^-(x)) \nonumber \\
&=&\sqrt{2} \Lambda \sigma \sigma_f \frac{n r^{n-1} \rho^n}{r^{2n} + \rho^{2n}}
\exp\left(i \sigma\left[(n + 1) \chi + \gamma + 
\sigma_f \frac{\pi}{4}\right]\right),
\nonumber \\
H^f_{--}&=&- \frac{1}{2 M'} \left[\p_i - i v^3_i(x)\right]^2 - 
\frac{n^2}{2 {\cal D}(\rho) \rho^2} 
\left(\p_\gamma + i n \frac{\Theta}{2 \pi} + 
i \sigma \frac{\rho^{2n}}{r^{2n} + \rho^{2n}}\right)^2 \nonumber \\
&=&- \frac{1}{2 M'} \left[\p_r^2 + \frac{1}{r} \p_r +
\frac{1}{r^2} \left(\p_\chi - i \sigma \frac{n \rho^{2n}}{r^{2n} + \rho^{2n}}
\right)^2 \right] \nonumber \\
&-&\frac{n^2}{2 {\cal D}(\rho) \rho^2} \left(\p_\gamma + 
i n \frac{\Theta}{2 \pi} + i \sigma \frac{\rho^{2n}}{r^{2n} + \rho^{2n}}
\right)^2.
\end{eqnarray}
We now make the ansatz
\begin{equation}
\label{ansatz2}
\Psi^f_{\sigma,m_+,m_-,m}(r,\chi,\gamma)\!=\!\left(\begin{array}{c} 
\psi_{\sigma,m_+,m_-,m,+}(r) 
\exp\left(i \sigma \left[m_+ \chi - \sigma_f \frac{\pi}{8}\right]\right) 
\exp(i \sigma (m - \frac{1}{2}) \gamma) \\ 
\sigma \sigma_f \psi_{\sigma,m_+,m_-,m,-}(r) 
\exp\left(i \sigma \left[m_- \chi + \sigma_f \frac{\pi}{8}\right]\right) 
\exp(i \sigma (m + \frac{1}{2}) \gamma) \\ \end{array} \right)
\end{equation}
with $m_- - m_+ = n + 1$. In order to ensure $2 \pi$-periodicity of the wave
function in the variable $\gamma$, $m$ must now be one half of some odd 
integer. This is in contrast to the rotating Skyrmion without a hole that was 
discussed in Subsection 2.4, for which $m$ was an integer. The radial 
Schr\"odinger equation is then given by
\begin{eqnarray}
H_r \psi_{\sigma,m_+,m_-,m}(r)&=&\left(\begin{array}{cc} 
H_{r++} & H_{r+-} \\ H_{r-+} & H_{r--} \end{array} \right)
\left(\begin{array}{c} \psi_{\sigma,m_+,m_-,m,+}(r) \\ \psi_{\sigma,m_+,m_-,m,-}(r) 
\end{array} \right) \nonumber \\
&=&E_{\sigma,m_+,m_-,m} \psi_{\sigma,m_+,m_-,m}(r).
\end{eqnarray}
In this case, the four matrix elements of the radial Hamiltonian $H_r$ take the 
form 
\begin{eqnarray}
\label{radialeq}
H_{r++}&=&- \frac{1}{2 M'} \left[\p_r^2 + \frac{1}{r} \p_r -
\frac{1}{r^2} \left(m_+ + \frac{n \rho^{2n}}{r^{2n} + \rho^{2n}}\right)^2\right]
\nonumber \\
&+&\frac{n^2}{2 {\cal D}(\rho) \rho^2} 
\left(m + \sigma n \frac{\Theta}{2 \pi} - \frac{1}{2} - 
\frac{\rho^{2n}}{r^{2n} + \rho^{2n}}\right)^2, \nonumber \\
H_{r+-}&=&H_{r-+} = 
\sqrt{2} \Lambda \frac{n r^{n-1} \rho^n}{r^{2n} + \rho^{2n}}, \nonumber \\
H_{r--}&=&- \frac{1}{2 M'} \left[\p_r^2 + \frac{1}{r} \p_r - 
\frac{1}{r^2} \left(m_- - \frac{n \rho^{2n}}{r^{2n} + \rho^{2n}}\right)^2\right]
\nonumber \\
&+&\frac{n^2}{2 {\cal D}(\rho) \rho^2} 
\left(m + \sigma n \frac{\Theta}{2 \pi} + \frac{1}{2} +
\frac{\rho^{2n}}{r^{2n} + \rho^{2n}}\right)^2.
\end{eqnarray}

\subsection{Symmetry Properties of a Single Hole Localized on a Skyrmion}

Let us again consider the spin operator (which generates an internal symmetry 
and is thus analogous to isospin in particle physics)
\begin{equation}
I = \left(\begin{array}{cc} - i \sigma \p_\gamma + \sigma n \frac{\Theta}{2 \pi} 
+ \frac{1}{2} & 0 \\ 0 & - i \sigma \p_\gamma + \sigma n \frac{\Theta}{2 \pi} - 
\frac{1}{2} \end{array}\right),
\end{equation}
which commutes with the Hamiltonian, i.e.\ $[H^f,I] = 0$. The wave function
$\Psi^f_{\sigma,m_+,m_-,m}$ is indeed an eigenstate of $I$, i.e.
\begin{equation}
I \ \Psi^f_{\sigma,m_+,m_-,m}(r,\chi,\gamma) =
\left(m + \sigma n \frac{\Theta}{2 \pi}\right) 
\Psi^f_{\sigma,m_+,m_-,m}(r,\chi,\gamma).
\end{equation}
Since $m$ is half of an odd integer, the rotating Skyrmion with one hole
localized on it has half-integer spin (or ``isospin''), at least for vanishing
anyon statistics parameter $\Theta = 0$.

The various symmetries such as the displacements $D_1'$ and $D_2'$, the $90$ 
degrees rotation $O$, as well as the reflection $R$, act on the wave function 
\begin{equation}
\Psi^f_{\sigma,n}(r,\chi,\gamma) = \left(\begin{array}{c}
\Psi^f_{\sigma,n,+}(r,\chi,\gamma) \\ \Psi^f_{\sigma,n,-}(r,\chi,\gamma)
\end{array} \right),
\end{equation}
of a single hole localized on a rotating (anti-)Skyrmion with winding number
$\sigma n$ as follows
\begin{eqnarray}
&&^{D_i'}\Psi^f_{\sigma,n}(r,\chi,\gamma) = \exp(i k^f_i a)
\left(\begin{array}{c}
\Psi^f_{\sigma,n,-}(r,\chi,\gamma) \\ - \Psi^f_{\sigma,n,+}(r,\chi,\gamma)
\end{array} \right), \nonumber \\
&&^O\Psi^f_{\sigma,n}(r,\chi,\gamma) = 
\left(\begin{array}{c}
\sigma_f \Psi^f_{\sigma,n,+}(r,\chi + \frac{\pi}{2},\gamma - n \frac{\pi}{2}) \\ 
\Psi^f_{\sigma,n,-}(r,\chi + \frac{\pi}{2},\gamma - n \frac{\pi}{2}) 
\end{array} \right), \nonumber \\
&&^R\Psi^f_{\sigma,n}(r,\chi,\gamma) = 
\left(\begin{array}{c}
\Psi^f_{\sigma,n,+}(r,- \chi,- \gamma) \\ 
\Psi^f_{\sigma,n,-}(r,- \chi,- \gamma) \end{array} \right).
\end{eqnarray}
For energy eigenstates this then implies
\begin{eqnarray}
&&^{D_i'}\Psi^f_{\sigma,m_+,m_-,m}(r,\chi,\gamma) = \sigma \sigma_f \exp(i k^f_i a) 
\Psi^f_{-\sigma,-m_-,-m_+,-m}(r,\chi,\gamma), \nonumber \\
&&^O\Psi^\alpha_{\sigma,m_+,m_-,m}(r,\chi,\gamma) = 
\exp\left(i \sigma [m_+ + m_- - 2 - 2 n m] \frac{\pi}{4} \right) 
\Psi^\beta_{\sigma,m_+,m_-,m}(r,\chi,\gamma), \nonumber \\
&&^O\Psi^\beta_{\sigma,m_+,m_-,m}(r,\chi,\gamma) = 
- \exp\left(i \sigma [m_+ + m_- - 2 n m] \frac{\pi}{4} \right) 
\Psi^\alpha_{\sigma,m_+,m_-,m}(r,\chi,\gamma), \nonumber \\
&&^R\Psi^\alpha_{\sigma,m_+,m_-,m}(r,\chi,\gamma) = 
\Psi^\beta_{-\sigma,m_+,m_-,m}(r,\chi,\gamma), \nonumber \\
&&^R\Psi^\beta_{\sigma,m_+,m_-,m}(r,\chi,\gamma) = 
\Psi^\alpha_{-\sigma,m_+,m_-,m}(r,\chi,\gamma).
\end{eqnarray}
It should be noted that for $\Theta \neq 0$ or $\pi$, the reflection symmetry 
$R$ is explicitly broken by the Hopf term. Assuming appropriate phase 
conventions for the radial wave functions, in the considerations of the shift 
symmetries $D_i'$, we have used
\begin{equation}
\psi_{-m_-,-m_+,-m,+}(r) = \psi_{m_+,m_-,m,-}(r), \quad
\psi_{-m_-,-m_+,-m,-}(r) = \psi_{m_+,m_-,m,+}(r),
\end{equation}
which follows from the behavior of eq.(\ref{radialeq}) under the replacement
of $m_+ \rightarrow m_+' = - m_-$, $m_- \rightarrow m_-' = - m_+$, and 
$m \rightarrow m' = - m$. It is worth noting that after this replacement the
constraint
\begin{equation}
m_-' - m_+' = - m_+ + m_- = n + 1
\end{equation}
remains satisfied.
 
\subsection{Schr\"odinger Equation for a Pair of Holes of Different Flavor
Localized on a Rotating Skyr\-mi\-on}

Let us now consider bound states of two holes localized on the same Skyrmion.
Both a hole of flavor $\alpha$ and another hole of flavor $\beta$ can occupy
the same single-particle ground state in a Skyrmion. For holes of the same 
flavor this would be forbidden by the Pauli principle. Since we are most
interested in the lowest energy states, we consider two holes of different
flavor. The case of two holes with the same flavor is discussed in Appendix A.
The Hamiltonian for two holes of different flavor $\alpha$ and $\beta$ is given 
by 
\begin{equation}
H = H^\alpha + H^\beta + H^\gamma,
\end{equation} 
where $H^\alpha$ and $H^\beta$ are the Hamiltonians for a hole of flavor 
$\alpha$ and $\beta$, respectively. Explicitly one has
\begin{eqnarray}
H^\alpha&=&\left(\begin{array}{cccc}  
H^\alpha_{++} & 0 & H^\alpha_{+-} & 0 \\
0 & H^\alpha_{++} & 0 & H^\alpha_{+-} \\
H^\alpha_{-+} & 0 & H^\alpha_{--} & 0 \\
0 & H^\alpha_{-+} & 0 & H^\alpha_{--} \end{array} \right), \quad
H^\beta = \left(\begin{array}{cccc}  
H^\beta_{++} & H^\beta_{+-} & 0 & 0 \\
H^\beta_{-+} & H^\beta_{--} & 0 & 0 \\
0 & 0 & H^\beta_{++} & H^\beta_{+-} \\
0 & 0 & H^\beta_{-+} & H^\beta_{--} \end{array} \right), \nonumber \\
H^\gamma&=&\left(\begin{array}{cccc}  
H^\gamma_{++++} & 0 & 0 & 0 \\
0 & H^\gamma_{+-+-} & 0 & 0 \\
0 & 0 & H^\gamma_{-+-+} & 0 \\
0 & 0 & 0 & H^\gamma_{----} \end{array} \right),
\end{eqnarray}
with
\begin{eqnarray}
\label{flavorHamiltonians}
&&H^\alpha_{++} = - \frac{1}{2 M'} (\p_i + i v^3_i(x))^2, \quad 
H^\alpha_{+-} = \Lambda (v_1^+(x) + v_2^+(x)), \nonumber \\
&&H^\alpha_{--} = - \frac{1}{2 M'} (\p_i - i v^3_i(x))^2, \quad
H^\alpha_{-+} = \Lambda (v_1^-(x) + v_2^-(x)), \nonumber \\
&&H^\beta_{++} = - \frac{1}{2 M'} (\p_i + i v^3_i(x))^2, \quad
H^\beta_{+-} = \Lambda (v_1^+(x) - v_2^+(x)), \nonumber \\
&&H^\beta_{--} = - \frac{1}{2 M'} (\p_i - i v^3_i(x))^2, \quad
H^\beta_{-+} = \Lambda (v_1^-(x) - v_2^-(x)), \nonumber \\
&&H^\gamma_{++++} = - \frac{n^2}{2 {\cal D}(\rho) \rho^2} 
\left(\p_\gamma + i n \frac{\Theta}{2 \pi}
- i \sigma \frac{\rho^{2n}}{r_\alpha^{2n} + \rho^{2n}}
- i \sigma \frac{\rho^{2n}}{r_\beta^{2n} + \rho^{2n}}\right)^2, \nonumber \\
&&H^\gamma_{+-+-} = - \frac{n^2}{2 {\cal D}(\rho) \rho^2} 
\left(\p_\gamma + i n \frac{\Theta}{2 \pi}
- i \sigma \frac{\rho^{2n}}{r_\alpha^{2n} + \rho^{2n}}
+ i \sigma \frac{\rho^{2n}}{r_\beta^{2n} + \rho^{2n}}\right)^2, \nonumber \\
&&H^\gamma_{-+-+} = - \frac{n^2}{2 {\cal D}(\rho) \rho^2} 
\left(\p_\gamma + i n \frac{\Theta}{2 \pi}
+ i \sigma \frac{\rho^{2n}}{r_\alpha^{2n} + \rho^{2n}}
- i \sigma \frac{\rho^{2n}}{r_\beta^{2n} + \rho^{2n}}\right)^2, \nonumber \\
&&H^\gamma_{----} = - \frac{n^2}{2 {\cal D}(\rho) \rho^2} 
\left(\p_\gamma + i n \frac{\Theta}{2 \pi}
+ i \sigma \frac{\rho^{2n}}{r_\alpha^{2n} + \rho^{2n}}
+ i \sigma \frac{\rho^{2n}}{r_\beta^{2n} + \rho^{2n}}\right)^2.
\end{eqnarray}
We now make the following ansatz for a two-hole energy eigenstate 
\begin{eqnarray}
&&\hskip-1.5cm
\Psi^{\alpha\beta}_{\sigma,m^\alpha_+,m^\alpha_-,m^\beta_+,m^\beta_-,m}
(r_\alpha,\chi_\alpha,r_\beta,\chi_\beta,\gamma) = \nonumber \\
&&\hskip-1.5cm\left(\begin{array}{c}
\psi_{\sigma,m^\alpha_+,m^\alpha_-,m^\beta_+,m^\beta_-,m,++}(r_\alpha,r_\beta)
\exp\left(i \sigma \left[m^\alpha_+ \chi_\alpha + m^\beta_+ \chi_\beta\right]
\right) \exp(i \sigma (m - 1) \gamma) \\
- \sigma 
\psi_{\sigma,m^\alpha_+,m^\alpha_-,m^\beta_+,m^\beta_-,m,+-}(r_\alpha,r_\beta)
\exp\left(i \sigma \left[m^\alpha_+ \chi_\alpha + m^\beta_- \chi_\beta - 
\frac{\pi}{4}\right]\right) \exp(i \sigma m \gamma) \\
\sigma
\psi_{\sigma,m^\alpha_+,m^\alpha_-,m^\beta_+,m^\beta_-,m,-+}(r_\alpha,r_\beta)
\exp\left(i \sigma \left[m^\alpha_- \chi_\alpha + m^\beta_+ \chi_\beta +
\frac{\pi}{4}\right] \right) \exp(i \sigma m \gamma) \\
- \psi_{\sigma,m^\alpha_+,m^\alpha_-,m^\beta_+,m^\beta_-,m,--}(r_\alpha,r_\beta)
\exp\left(i \sigma \left[m^\alpha_- \chi_\alpha + m^\beta_- \chi_\beta\right]
\right) \exp(i \sigma (m + 1) \gamma) \end{array}\right). \nonumber \\ \
\end{eqnarray}
Again, this solves the Schr\"odinger equation only if $m^f_- - m^f_+ = n + 1$.
As for the Skyrmion without holes, in this case, $m$ is again an integer. The 
resulting radial Schr\"odinger equation then takes the form
\begin{equation}
\label{twoholeradial}
H_r \psi_{\sigma,m^\alpha_+,m^\alpha_-,m^\beta_+,m^\beta_-,m}(r_\alpha,r_\beta) =
E_{\sigma,m^\alpha_+,m^\alpha_-,m^\beta_+,m^\beta_-,m} 
\psi_{\sigma,m^\alpha_+,m^\alpha_-,m^\beta_+,m^\beta_-,m}(r_\alpha,r_\beta),
\end{equation}
with
\begin{equation}
\psi_{\sigma,m^\alpha_+,m^\alpha_-,m^\beta_+,m^\beta_-,m}(r_\alpha,r_\beta) =
\left(\begin{array}{c}
\psi_{\sigma,m^\alpha_+,m^\alpha_-,m^\beta_+,m^\beta_-,m,++}(r_\alpha,r_\beta) \\
\psi_{\sigma,m^\alpha_+,m^\alpha_-,m^\beta_+,m^\beta_-,m,+-}(r_\alpha,r_\beta) \\
\psi_{\sigma,m^\alpha_+,m^\alpha_-,m^\beta_+,m^\beta_-,m,-+}(r_\alpha,r_\beta) \\
\psi_{\sigma,m^\alpha_+,m^\alpha_-,m^\beta_+,m^\beta_-,m,--}(r_\alpha,r_\beta) \\
\end{array}\right).
\end{equation}
The radial Hamiltonian is given by
\begin{equation}
H_r = H_r^\alpha + H_r^\beta + H_r^\gamma,
\end{equation} 
with
\begin{eqnarray}
H_r^\alpha&=&\left(\begin{array}{cccc}  
H^\alpha_{r++} & 0 & H^\alpha_{r+-} & 0 \\
0 & H^\alpha_{r++} & 0 & H^\alpha_{r+-} \\
H^\alpha_{r-+} & 0 & H^\alpha_{r--} & 0 \\
0 & H^\alpha_{r-+} & 0 & H^\alpha_{r--} \end{array} \right), \nonumber \\
H_r^\beta&=&\left(\begin{array}{cccc}  
H^\beta_{r++} & H^\beta_{r+-} & 0 & 0 \\
H^\beta_{r-+} & H^\beta_{r--} & 0 & 0 \\
0 & 0 & H^\beta_{r++} & H^\beta_{r+-} \\
0 & 0 & H^\beta_{r-+} & H^\beta_{r--} \end{array} \right), \nonumber \\
H_r^\gamma&=&\left(\begin{array}{cccc}  
H^\gamma_{r++++} & 0 & 0 & 0 \\
0 & H^\gamma_{r+-+-} & 0 & 0 \\
0 & 0 & H^\gamma_{r-+-+} & 0 \\
0 & 0 & 0 & H^\gamma_{r----} \end{array} \right).
\end{eqnarray}
The matrix elements of the fermionic part of the radial Hamiltonian are
\begin{eqnarray}
H^f_{r++}&=&- \frac{1}{2 M'} 
\left[\p_{r_f}^2 + \frac{1}{r_f} \p_{r_f} - \frac{1}{r_f^2} 
\left(m^f_+ + \frac{n \rho^{2n}}{r_f^{2n} + \rho^{2n}}\right)^2\right], 
\nonumber \\
H^f_{r+-}&=&H^f_{r-+} = 
\sqrt{2} \Lambda \frac{n r_f^{n-1} \rho^n}{r_f^{2n} + \rho^{2n}}, \nonumber \\
H^f_{r--}&=&- \frac{1}{2 M'} 
\left[\p_{r_f}^2 + \frac{1}{r_f} \p_{r_f} - \frac{1}{r_f^2} 
\left(m^f_- - \frac{n \rho^{2n}}{r_f^{2n} + \rho^{2n}}\right)^2\right],
\end{eqnarray}
while the rotational Skyrmion contributions are given by
\begin{eqnarray}
H^\gamma_{r++++}&=&\frac{n^2}{2 {\cal D}(\rho) \rho^2} 
\left(m + \sigma n \frac{\Theta}{2 \pi} - 1 -  
\frac{\rho^{2n}}{r_\alpha^{2n} + \rho^{2n}} -
\frac{\rho^{2n}}{r_\beta^{2n} + \rho^{2n}}\right)^2, \nonumber \\
H^\gamma_{r+-+-}&=&\frac{n^2}{2 {\cal D}(\rho) \rho^2} 
\left(m + \sigma n \frac{\Theta}{2 \pi} -  
\frac{\rho^{2n}}{r_\alpha^{2n} + \rho^{2n}} +
\frac{\rho^{2n}}{r_\beta^{2n} + \rho^{2n}}\right)^2, \nonumber \\
H^\gamma_{r-+-+}&=&\frac{n^2}{2 {\cal D}(\rho) \rho^2} 
\left(m + \sigma n \frac{\Theta}{2 \pi} +  
\frac{\rho^{2n}}{r_\alpha^{2n} + \rho^{2n}} -
\frac{\rho^{2n}}{r_\beta^{2n} + \rho^{2n}}\right)^2, \nonumber \\
H^\gamma_{r----}&=&\frac{n^2}{2 {\cal D}(\rho) \rho^2} 
\left(m + \sigma n \frac{\Theta}{2 \pi} + 1 +
\frac{\rho^{2n}}{r_\alpha^{2n} + \rho^{2n}} +
\frac{\rho^{2n}}{r_\beta^{2n} + \rho^{2n}}\right)^2.
\end{eqnarray}

\subsection{Symmetry Properties of a Pair of Holes with Different Flavors
Localized on a Skyrmion}

It is worth noticing that the spin operator $I$, which commutes with the 
two-hole Hamiltonian $H$, is given by
\begin{equation}
\label{spinoperator}
I = \left(\begin{array}{cccc}  
- i \sigma \p_\gamma + \sigma n \frac{\Theta}{2 \pi} + 1 & 0 & 0 & 0 \\
0 & - i \sigma \p_\gamma + \sigma n \frac{\Theta}{2 \pi} & 0 & 0 \\
0 & 0 & - i \sigma \p_\gamma + \sigma n \frac{\Theta}{2 \pi} & 0 \\
0 & 0 & 0 & - i \sigma \p_\gamma + \sigma n \frac{\Theta}{2 \pi} - 1 
\end{array} \right),
\end{equation}
such that 
\begin{eqnarray}
&&I \Psi^{\alpha\beta}_{\sigma,m^\alpha_+,m^\alpha_-,m^\beta_+,m^\beta_-,m}
(r_\alpha,\chi_\alpha,r_\beta,\chi_\beta,\gamma) = \nonumber \\
&&\left(m + \sigma n \frac{\Theta}{2 \pi}\right) 
\Psi^{\alpha\beta}_{\sigma,m^\alpha_+,m^\alpha_-,m^\beta_+,m^\beta_-,m}
(r_\alpha,\chi_\alpha,r_\beta,\chi_\beta,\gamma).
\end{eqnarray}
Since $m$ is an integer, as expected, for $\Theta = 0$ the state with two
holes localized on a Skyrmion has integer spin (which plays the role of
``isospin'').

The symmetries $D_i'$, $O$, and $R$ act on a general two-hole wave function
\begin{equation}
\Psi^{\alpha \beta}_{\sigma,n}(r_\alpha,\chi_\alpha,r_\beta,\chi_\beta,\gamma) =
\left(\begin{array}{c} 
\Psi^{\alpha \beta}_{\sigma,n,++}(r_\alpha,\chi_\alpha,r_\beta,\chi_\beta,\gamma) \\
\Psi^{\alpha \beta}_{\sigma,n,+-}(r_\alpha,\chi_\alpha,r_\beta,\chi_\beta,\gamma) \\
\Psi^{\alpha \beta}_{\sigma,n,-+}(r_\alpha,\chi_\alpha,r_\beta,\chi_\beta,\gamma) \\
\Psi^{\alpha \beta}_{\sigma,n,--}(r_\alpha,\chi_\alpha,r_\beta,\chi_\beta,\gamma)
\end{array}\right)
\end{equation}
as follows
\begin{eqnarray}
&&^{D_i'}\Psi^{\alpha \beta}_{\sigma,n}(r_\alpha,\chi_\alpha,r_\beta,\chi_\beta,\gamma) =
\exp(i(k_i^\alpha + k_i^\beta) a) \left(\begin{array}{c} 
\Psi^{\alpha \beta}_{\sigma,n,--}(r_\alpha,\chi_\alpha,r_\beta,\chi_\beta,\gamma) \\
- \Psi^{\alpha \beta}_{\sigma,n,-+}(r_\alpha,\chi_\alpha,r_\beta,\chi_\beta,\gamma) \\
- \Psi^{\alpha \beta}_{\sigma,n,+-}(r_\alpha,\chi_\alpha,r_\beta,\chi_\beta,\gamma) \\
\Psi^{\alpha \beta}_{\sigma,n,++}(r_\alpha,\chi_\alpha,r_\beta,\chi_\beta,\gamma)
\end{array}\right), \nonumber \\
&&^O\Psi^{\alpha \beta}_{\sigma,n}(r_\alpha,\chi_\alpha,r_\beta,\chi_\beta,\gamma) =
\left(\begin{array}{c} 
- \Psi^{\alpha \beta}_{\sigma,n,++}(r_\beta,\chi_\beta + \frac{\pi}{2},
r_\alpha,\chi_\alpha + \frac{\pi}{2},\gamma - n \frac{\pi}{2}) \\
- \Psi^{\alpha \beta}_{\sigma,n,-+}(r_\beta,\chi_\beta + \frac{\pi}{2},
r_\alpha,\chi_\alpha + \frac{\pi}{2},\gamma - n \frac{\pi}{2}) \\
\Psi^{\alpha \beta}_{\sigma,n,+-}(r_\beta,\chi_\beta + \frac{\pi}{2},
r_\alpha,\chi_\alpha + \frac{\pi}{2},\gamma - n \frac{\pi}{2}) \\
\Psi^{\alpha \beta}_{\sigma,n,--}(r_\beta,\chi_\beta + \frac{\pi}{2},
r_\alpha,\chi_\alpha + \frac{\pi}{2},\gamma - n \frac{\pi}{2})
\end{array}\right), \nonumber \\
&&^R\Psi^{\alpha \beta}_{\sigma,n}(r_\alpha,\chi_\alpha,r_\beta,\chi_\beta,\gamma) =
\left(\begin{array}{c} 
\Psi^{\alpha \beta}_{\sigma,n,++}(r_\beta,-\chi_\beta,r_\alpha,-\chi_\alpha,-\gamma) \\
\Psi^{\alpha \beta}_{\sigma,n,-+}(r_\beta,-\chi_\beta,r_\alpha,-\chi_\alpha,-\gamma) \\
\Psi^{\alpha \beta}_{\sigma,n,+-}(r_\beta,-\chi_\beta,r_\alpha,-\chi_\alpha,-\gamma) \\
\Psi^{\alpha \beta}_{\sigma,n,--}(r_\beta,-\chi_\beta,r_\alpha,-\chi_\alpha,-\gamma)
\end{array}\right).
\end{eqnarray}

It is straightforward to show that for the two-hole energy eigenstates this
implies 
\begin{eqnarray}
\label{symtwoholes}
^{D_1'}\Psi^{\alpha\beta}_{\sigma,m^\alpha_+,m^\alpha_-,m^\beta_+,m^\beta_-,m}
(r_\alpha,\chi_\alpha,r_\beta,\chi_\beta,\gamma)\!\!\!\!&=&\!\!\!\!
\Psi^{\alpha\beta}_{-\sigma,-m^\alpha_-,-m^\alpha_+,-m^\beta_-,-m^\beta_+,-m}
(r_\alpha,\chi_\alpha,r_\beta,\chi_\beta,\gamma), \nonumber \\
^{D_2'}\Psi^{\alpha\beta}_{\sigma,m^\alpha_+,m^\alpha_-,m^\beta_+,m^\beta_-,m}
(r_\alpha,\chi_\alpha,r_\beta,\chi_\beta,\gamma)\!\!\!\!&=&\!\!\!\!
- \Psi^{\alpha\beta}_{-\sigma,-m^\alpha_-,-m^\alpha_+,-m^\beta_-,-m^\beta_+,-m}
(r_\alpha,\chi_\alpha,r_\beta,\chi_\beta,\gamma), \nonumber \\
^O\Psi^{\alpha\beta}_{\sigma,m^\alpha_+,m^\alpha_-,m^\beta_+,m^\beta_-,m}
(r_\alpha,\chi_\alpha,r_\beta,\chi_\beta,\gamma)\!\!\!\!&=&\!\!\!\!
\exp\left(i \sigma [m^\alpha_+ + m^\beta_- - m n + 1] \frac{\pi}{2}\right)
\nonumber \\
&\times&\!\!\!\!\Psi^{\alpha\beta}_{\sigma,m^\alpha_+,m^\alpha_-,m^\beta_+,m^\beta_-,m}
(r_\alpha,\chi_\alpha,r_\beta,\chi_\beta,\gamma), \nonumber \\
^R\Psi^{\alpha\beta}_{\sigma,m^\alpha_+,m^\alpha_-,m^\beta_+,m^\beta_-,m}
(r_\alpha,\chi_\alpha,r_\beta,\chi_\beta,\gamma)\!\!\!\!&=&\!\!\!\!
\Psi^{\alpha\beta}_{-\sigma,m^\beta_+,m^\beta_-,m^\alpha_+,m^\alpha_-,m}
(r_\alpha,\chi_\alpha,r_\beta,\chi_\beta,\gamma).
\end{eqnarray}
Here we have assumed an appropriate phase convention for the radial wave 
function $\psi_{\sigma,m^\alpha_+,m^\alpha_-,m^\beta_+,m^\beta_-,m}(r_\alpha,r_\beta)$. In the 
context of the shift symmetries $D_i'$ we have used
\begin{eqnarray}
&&\psi_{\sigma,m^\alpha_+,m^\alpha_-,m^\beta_+,m^\beta_-,m,--}(r_\alpha,r_\beta) = 
\psi_{-\sigma,-m^\alpha_-,-m^\alpha_+,-m^\beta_-,-m^\beta_+,-m,++}(r_\alpha,r_\beta), \nonumber \\
&&\psi_{\sigma,m^\alpha_+,m^\alpha_-,m^\beta_+,m^\beta_-,m,-+}(r_\alpha,r_\beta) = 
\psi_{-\sigma,-m^\alpha_-,-m^\alpha_+,-m^\beta_-,-m^\beta_+,-m,+-}(r_\alpha,r_\beta), \nonumber \\
&&\psi_{\sigma,m^\alpha_+,m^\alpha_-,m^\beta_+,m^\beta_-,m,+-}(r_\alpha,r_\beta) = 
\psi_{-\sigma,-m^\alpha_-,-m^\alpha_+,-m^\beta_-,-m^\beta_+,-m,-+}(r_\alpha,r_\beta), \nonumber \\
&&\psi_{\sigma,m^\alpha_+,m^\alpha_-,m^\beta_+,m^\beta_-,m,++}(r_\alpha,r_\beta) = 
\psi_{-\sigma,-m^\alpha_-,-m^\alpha_+,-m^\beta_-,-m^\beta_+,-m,--}(r_\alpha,r_\beta).
\end{eqnarray}
These relations follow from the symmetries of the radial Schr\"odinger equation
(\ref{twoholeradial}). Similarly, in the context of the rotation $O$ we have 
used 
\begin{eqnarray}
&&\psi_{\sigma,m^\alpha_+,m^\alpha_-,m^\beta_+,m^\beta_-,m,++}(r_\beta,r_\alpha) = 
\psi_{\sigma,m^\beta_+,m^\beta_-,m^\alpha_+,m^\alpha_-,m,++}(r_\alpha,r_\beta), \nonumber \\
&&\psi_{\sigma,m^\alpha_+,m^\alpha_-,m^\beta_+,m^\beta_-,m,-+}(r_\beta,r_\alpha) = 
\psi_{\sigma,m^\beta_+,m^\beta_-,m^\alpha_+,m^\alpha_-,m,+-}(r_\alpha,r_\beta), \nonumber \\
&&\psi_{\sigma,m^\alpha_+,m^\alpha_-,m^\beta_+,m^\beta_-,m,+-}(r_\beta,r_\alpha) = 
\psi_{\sigma,m^\beta_+,m^\beta_-,m^\alpha_+,m^\alpha_-,m,-+}(r_\alpha,r_\beta), \nonumber \\
&&\psi_{\sigma,m^\alpha_+,m^\alpha_-,m^\beta_+,m^\beta_-,m,--}(r_\beta,r_\alpha) = 
\psi_{\sigma,m^\beta_+,m^\beta_-,m^\alpha_+,m^\alpha_-,m,--}(r_\alpha,r_\beta).
\end{eqnarray}
Finally, in the context of the reflection symmetry $R$ we have used
\begin{eqnarray}
\label{Rsym}
&&\psi_{\sigma,m^\alpha_+,m^\alpha_-,m^\beta_+,m^\beta_-,m,++}(r_\beta,r_\alpha) = 
\psi_{-\sigma,m^\beta_+,m^\beta_-,m^\alpha_+,m^\alpha_-,m,++}(r_\alpha,r_\beta), \nonumber \\
&&\psi_{\sigma,m^\alpha_+,m^\alpha_-,m^\beta_+,m^\beta_-,m,-+}(r_\beta,r_\alpha) = 
\psi_{-\sigma,m^\beta_+,m^\beta_-,m^\alpha_+,m^\alpha_-,m,+-}(r_\alpha,r_\beta), \nonumber \\
&&\psi_{\sigma,m^\alpha_+,m^\alpha_-,m^\beta_+,m^\beta_-,m,+-}(r_\beta,r_\alpha) = 
\psi_{-\sigma,m^\beta_+,m^\beta_-,m^\alpha_+,m^\alpha_-,m,-+}(r_\alpha,r_\beta), \nonumber \\
&&\psi_{\sigma,m^\alpha_+,m^\alpha_-,m^\beta_+,m^\beta_-,m,--}(r_\beta,r_\alpha) = 
\psi_{-\sigma,m^\beta_+,m^\beta_-,m^\alpha_+,m^\alpha_-,m,--}(r_\alpha,r_\beta).
\end{eqnarray}
The relations in eq.(\ref{Rsym}) follow from the symmetries of the radial 
Schr\"odinger equation (\ref{twoholeradial}) for $\Theta = 0$. For 
$\Theta \neq 0$ or $\pi$, the Hopf term explicitly breaks the reflection 
symmetry.

\subsection{Comparison with Two-Hole States Bound by One-Mag\-non Exchange}

In \cite{Bru06} states of two holes bound by one-magnon exchange in a square
lattice antiferromagnet have been investigated in great detail. Here we 
summarize as well as extend some of the relevant results. In the rest frame, 
the Schr\"odinger equation for two holes of flavor $\alpha$ and $\beta$ takes 
the form
\begin{equation}
\left(\begin{array}{cc} - \frac{1}{M'} \Delta & V^{\alpha\beta}(\vec r) 
\\[0.2ex]
V^{\alpha\beta}(\vec r) &  - \frac{1}{M'} \Delta \end{array} \right)
\left(\begin{array}{c} \Psi_1(\vec r) \\ 
\Psi_2(\vec r) \end{array}\right) = E 
\left(\begin{array}{c} \Psi_1(\vec r) \\ 
\Psi_2(\vec r) \end{array}\right).
\end{equation}
The components $\Psi_1(\vec r)$ and $\Psi_2(\vec r)$ are probability
amplitudes for the spin-flavor combinations $\alpha_+\beta_-$ and 
$\alpha_-\beta_+$, respectively. The potential 
\begin{equation}
V^{\alpha\beta}(\vec r) =  
\frac{\Lambda^2}{2 \pi \rho_s} \frac{\cos(2 \varphi)}{r^2}
\end{equation}
couples the two channels because magnon exchange is accompanied by a spin-flip.
Here $\vec r = \vec r_+ - \vec r_-$ is the distance vector between the two 
holes of spin $+$ and $-$ and $\varphi$ is the angle between $\vec r$ and the 
$x$-axis. Magnon exchange is attractive between holes of opposite spin, and 
hence magnon-mediated two-hole bound states are invariant under the unbroken
subgroup $U(1)_s$. We make the ansatz 
\begin{equation}
\Psi_1(\vec r) \pm \Psi_2(\vec r) = R(r) \chi_\pm(\varphi).
\end{equation}
For the angular part of the wave function this implies
\begin{equation}
- \frac{d^2\chi_\pm(\varphi)}{d\varphi^2} \pm 
\frac{M' \Lambda^2}{2 \pi \rho_s} \cos(2 \varphi) \chi_\pm(\varphi) = 
- \lambda \chi_\pm(\varphi).
\end{equation}
This is a Mathieu equation whose solution with the lowest eigenvalue 
$- \lambda_1$ is given by
\begin{equation}
\chi^1_\pm(\varphi) = \frac{1}{\sqrt{\pi}}
\mbox{ce}_0 \left(\varphi,\pm \frac{M' \Lambda^2}{4 \pi \rho_s}\right), \quad
\lambda_1 = \frac{1}{2} \left(\frac{M' \Lambda^2}{4 \pi \rho_s}\right)^2 + 
{\cal O}(\Lambda^8).
\end{equation}
The first excited state and its eigenvalue $- \lambda_2$ is given by
\begin{eqnarray}
&&\chi^2_+(\varphi) = \frac{1}{\sqrt{\pi}}
\mbox{se}_1 \left(\varphi,\frac{M' \Lambda^2}{4 \pi \rho_s}\right), \nonumber \\
&&\chi^2_-(\varphi) = \frac{1}{\sqrt{\pi}}
\mbox{se}_1 \left(\varphi - \frac{\pi}{2},
\frac{M' \Lambda^2}{4 \pi \rho_s}\right) = - \frac{1}{\sqrt{\pi}}
\mbox{ce}_1 \left(\varphi,- \frac{M' \Lambda^2}{4 \pi \rho_s}\right), 
\nonumber \\
&&\lambda_2 = - 1 + \frac{M' \Lambda^2}{4 \pi \rho_s} + 
\frac{1}{8} \left(\frac{M' \Lambda^2}{4 \pi \rho_s}\right)^2 -
\frac{1}{64} \left(\frac{M' \Lambda^2}{4 \pi \rho_s}\right)^3 +
{\cal O}(\Lambda^8). 
\end{eqnarray}
For small $\Lambda$, $\lambda_2 < 0$, which (as we will see) implies that the 
corresponding two-hole state is unbound. For 
$M' \Lambda^2/4 \pi \rho_s > 0.908046$, on the other hand, 
$\lambda_1, \lambda_2 > 0$, such that then both states are bound. The periodic 
Mathieu functions 
$\mbox{ce}_0(\varphi,M' \Lambda^2/4 \pi \rho_s)$ and
$\mbox{se}_1 (\varphi,M' \Lambda^2/4 \pi \rho_s)$ \cite{Abr72} are shown in 
figure 3.
\begin{figure}[tb]
\begin{center}
\vspace{-0.3cm}
\epsfig{file=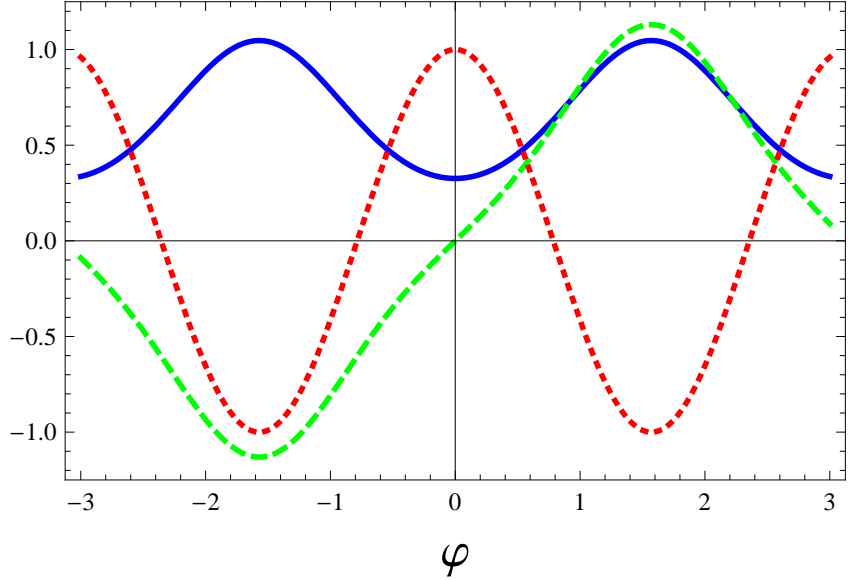,width=11cm}
\end{center}
\caption{\it Angular wave functions
$\mbox{ce}_0(\varphi,M' \Lambda^2/4 \pi \rho_s)$ (solid curve) and
$\mbox{se}_1(\varphi,M' \Lambda^2/4 \pi \rho_s)$ (dashed curve) as well as the
angle-dependence $\cos(2 \varphi)$ of the potential (dotted curve) for a pair 
of holes with flavors $\alpha$ and $\beta$ 
($M' \Lambda^2/4 \pi \rho_s = 1.25$).}
\end{figure}
The corresponding radial Schr\"odinger equation is given by
\begin{equation}
\label{radial}
- \left[\frac{d^2R_i(r)}{dr^2} + 
\frac{1}{r} \frac{dR_i(r)}{dr}\right] - \frac{\lambda_i}{r^2} R_i(r) = 
M' E_i R_i(r), \quad i \in \{1,2\}.
\end{equation}
The short-distance repulsion between two holes can be incorporated by a hard 
core of radius $r_0$, i.e.\ we require $R_i(r_0) = 0$. The radial Schr\"odinger 
equation for the bound states is solved by a Bessel function
\begin{equation}
R_i(r) = A_i K_\nu \big( \sqrt{M' |E_{ik}|} r \big), \quad k = 1,2,3,\dots, 
\quad \nu = i \sqrt{\lambda_i}.
\end{equation}
The energy (determined from $K_\nu \big( \sqrt{M' |E_{ik}|} r_0 \big) = 0$)
is then given by
\begin{equation}
E_{ik} \sim - (M' r_0^2)^{-1} \exp(- 2 \pi k/\sqrt{\lambda_i})
\end{equation}
for large $n$. Magnon exchange mediates weak attractive forces that lead to a
small binding energy.

The two lowest energy states with angular part $\chi^1_+(\varphi)$ and 
$\chi^1_-(\varphi)$ are degenerate in energy. Linearly combining the two states 
to two eigenstates of the rotation $O$, one obtains
\begin{equation}
\Psi^1_\pm(\vec r) = R_1(r) \left(\begin{array}{c}
\chi^1_+(\varphi) \mp i \chi^1_-(\varphi) \\
\chi^1_+(\varphi) \pm i \chi^1_-(\varphi) \end{array} \right).
\end{equation}
The corresponding probability density is illustrated in figure 4 (left panel).
\begin{figure}[tb]
\begin{center}
\vspace{-0.3cm}
\epsfig{file=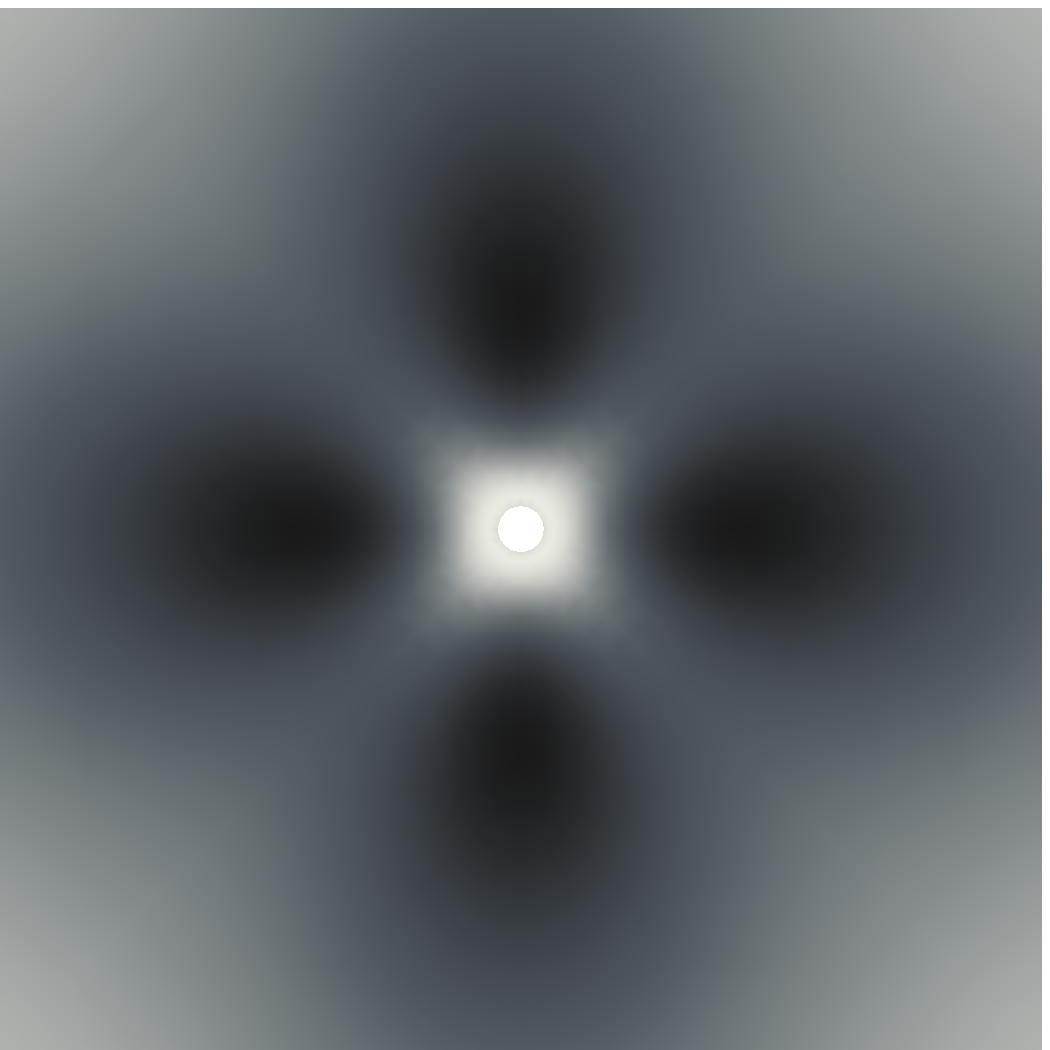,width=7.5cm}
\epsfig{file=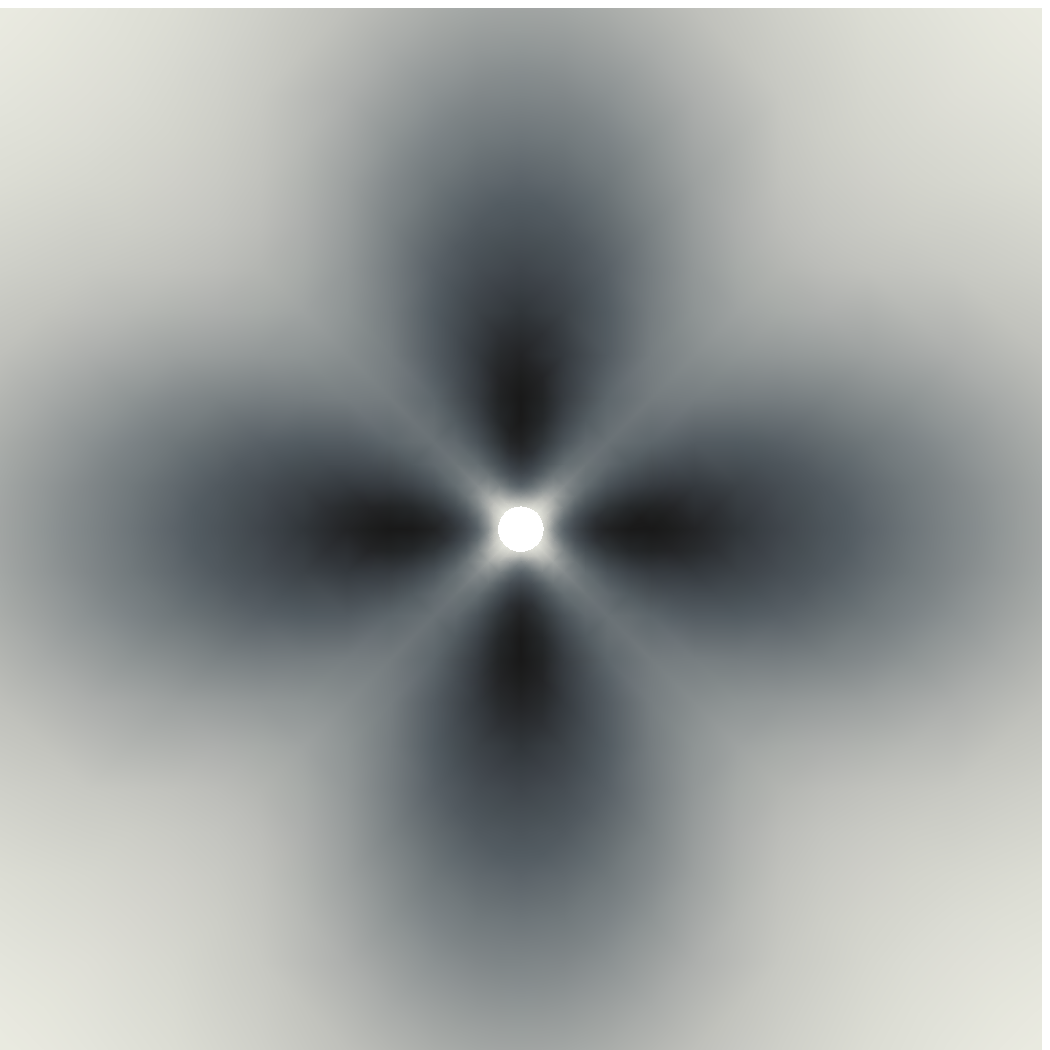,width=7.5cm}
\end{center}
\caption{\it Probability distribution for two holes with flavors $\alpha$ and 
$\beta$. Left panel: the ground state with p-wave symmetry. Right panel:
excited states with s- or d-wave symmetry, but with identical probability
densities ($M' \Lambda^2/4 \pi \rho_s = 1.25$, $r_0 = a$).}
\end{figure}
While the probability density seems to resemble $d_{x^2-y^2}$ symmetry, unlike 
for an actual d-wave, the wave function is suppressed, but not equal to zero, 
along the lattice diagonals. In fact, as one operates on the states 
$\Psi^1_\pm(\vec r)$ with the 90 degrees rotation $O$, one obtains the
eigenvalues $\pm i$, which shows that they actually have p-wave symmetry. 

Under the discrete symmetries $D_i'$, $O$, and $R$, the ground states 
$\Psi^1_\pm(\vec r)$, which are bound by magnon exchange, transform as
\begin{eqnarray}
^{D_1'}\Psi^1_\pm(\vec r)&=&R_1(r) \left(\begin{array}{c}
\chi^1_+(\varphi) \pm i \chi^1_-(\varphi) \\
\chi^1_+(\varphi) \mp i \chi^1_-(\varphi) \end{array} \right) = 
\Psi^1_\mp(\vec r),
\nonumber \\
^{D_2'}\Psi^1_\pm(\vec r)&=&- R_1(r) \left(\begin{array}{c}
\chi^1_+(\varphi) \pm i \chi^1_-(\varphi) \\
\chi^1_+(\varphi) \mp i \chi^1_-(\varphi) \end{array} \right) = 
- \Psi^1_\mp(\vec r), \nonumber \\
^O\Psi^1_\pm(\vec r)&=&R_1(r) \left(\begin{array}{c}
\chi^1_+(\varphi + \frac{\pi}{2}) \pm i \chi^1_-(\varphi + \frac{\pi}{2}) \\
- \chi^1_+(\varphi + \frac{\pi}{2}) \pm i \chi^1_-(\varphi + \frac{\pi}{2}) 
\end{array} \right) \nonumber \\
&=&R_1(r) \left(\begin{array}{c}
\chi^1_-(\varphi) \pm i \chi^1_+(\varphi) \\
- \chi^1_-(\varphi) \pm i \chi^1_+(\varphi) \end{array} \right) = 
\pm i \Psi^1_\pm(\vec r), \nonumber \\
^R\Psi^1_\pm(\vec r)&=&R_1(r) \left(\begin{array}{c}
\chi^1_+(- \varphi) \pm i \chi^1_-(- \varphi) \\
\chi^1_+(- \varphi) \mp i \chi^1_-(- \varphi) \end{array} \right) \nonumber \\
&=&R_1(r) \left(\begin{array}{c}
\chi^1_+(\varphi) \pm i \chi^1_-(\varphi) \\
\chi^1_+(\varphi) \mp i \chi^1_-(\varphi) \end{array} \right) = 
\Psi^1_\mp(\vec r).
\end{eqnarray}
It should be noted that in \cite{Bru06} there are two typos in the last line
of the previous equation for the reflection symmetry $R$ (eq.(6.20) in 
\cite{Bru06}).

Remarkably, the magnon-mediated two-hole ground states $\Psi^1_\pm(\vec r)$
transform exactly as the two-hole states localized on a rotating Skyrmion
with $n = 1$, provided that we associate $\Psi^1_\pm(\vec r)$ with the 
corresponding two-hole-Skyrmion wave function
$\Psi^{\alpha\beta}_{\pm,-1,1,-1,1,0}(r_\alpha,\chi_\alpha,r_\beta,\chi_\beta,\gamma)$ 
with the quantum numbers $\sigma = \pm$, $m^\alpha_+ = m^\beta_+ = -1$,
$m^\alpha_- = m^\beta_- = 1$, and $m = 0$. Indeed, according to 
eq.(\ref{symtwoholes}) one obtains
\begin{eqnarray}
^{D_1'}\Psi^{\alpha\beta}_{\pm,-1,1,-1,1,0}(r_\alpha,\chi_\alpha,r_\beta,\chi_\beta,\gamma)
&=&\Psi^{\alpha\beta}_{\mp,-1,1,-1,1,0}(r_\alpha,\chi_\alpha,r_\beta,\chi_\beta,\gamma), 
\nonumber \\
^{D_2'}\Psi^{\alpha\beta}_{\pm,-1,1,-1,1,0}(r_\alpha,\chi_\alpha,r_\beta,\chi_\beta,\gamma)
&=&- \Psi^{\alpha\beta}_{\mp,-1,1,-1,1,0}(r_\alpha,\chi_\alpha,r_\beta,\chi_\beta,\gamma),
\nonumber \\
^O\Psi^{\alpha\beta}_{\pm,-1,1,-1,1,0}(r_\alpha,\chi_\alpha,r_\beta,\chi_\beta,\gamma)
&=&\pm i 
\Psi^{\alpha\beta}_{\pm,-1,1,-1,1,0}(r_\alpha,\chi_\alpha,r_\beta,\chi_\beta,\gamma), 
\nonumber \\
^R\Psi^{\alpha\beta}_{\pm,-1,1,-1,1,0}(r_\alpha,\chi_\alpha,r_\beta,\chi_\beta,\gamma)
&=&\Psi^{\alpha\beta}_{\mp,-1,1,-1,1,0}(r_\alpha,\chi_\alpha,r_\beta,\chi_\beta,\gamma).
\end{eqnarray}
Just as the magnon-mediated bound 
states, these states are also invariant under $U(1)_s$ and they have fermion
number 2. One may argue that the two-hole-Skyrmion states, in addition, have
Skyrmion number as a conserved topological quantum number. However, as we 
discussed before, Skyrmion number has no analog in the underlying microscopic
Hubbard or $t$-$J$ models and is just an accidental symmetry of the effective 
theory. We thus conclude that, in their ground state, two holes bound by magnon 
exchange indeed have exactly the same quantum numbers as two holes localized on 
a rotating Skyrmion with $n = 1$. This implies that these sets of states may 
evolve into each other upon doping. In this way, two holes weakly bound by 
magnon exchange at small doping may evolve into a strongly correlated preformed 
pair of holes localized on a Skyrmion. However, as we have just seen, theses
bound states actually have p-wave symmetry.

Let us also consider the excited states, bound by magnon exchange, with angular 
part $\chi^2_+(\varphi)$ and $\chi^2_-(\varphi)$, which are again degenerate. 
Linearly combining these two states to two eigenstates of the rotation $O$, one 
obtains
\begin{equation}
\Psi^2_\pm(\vec r) = R_2(r) \left(\begin{array}{c}
\chi^2_+(\varphi) \mp \chi^2_-(\varphi) \\
\chi^2_+(\varphi) \pm \chi^2_-(\varphi) \end{array} \right).
\end{equation}
Operating on the states $\Psi^2_\pm(\vec r)$ with the 90 degrees rotation $O$, 
one now obtains the eigenvalues $\pm 1$, which implies that $\Psi^2_+(\vec r)$
represents an s-wave, while $\Psi^2_-(\vec r)$ actually has d-wave symmetry. As
a consequence of an interplay of the various symmetries, the two states are 
exactly degenerate. The corresponding probability density is illustrated in 
figure 4 (right panel). Interestingly, although the states have different
symmetries, their probability densities are identical.

Under the discrete symmetries $D_i'$, $O$, and $R$, the excited states 
$\Psi^2_\pm(\vec r)$, which are bound by magnon exchange, transform as
\begin{eqnarray}
^{D_1'}\Psi^2_\pm(\vec r)&=&- R_2(r) \left(\begin{array}{c}
\chi^2_+(\varphi) \pm \chi^2_-(\varphi) \\
\chi^2_+(\varphi) \mp \chi^2_-(\varphi) \end{array} \right) = 
- \Psi^2_\mp(\vec r), \nonumber \\
^{D_2'}\Psi^2_\pm(\vec r)&=&R_2(r) \left(\begin{array}{c}
\chi^2_+(\varphi) \pm \chi^2_-(\varphi) \\
\chi^2_+(\varphi) \mp \chi^2_-(\varphi) \end{array} \right) = 
\Psi^2_\mp(\vec r), \nonumber \\
^O\Psi^2_\pm(\vec r)&=&R_2(r) \left(\begin{array}{c}
\chi^2_+(\varphi + \frac{\pi}{2}) \pm \chi^2_-(\varphi + \frac{\pi}{2}) \\
- \chi^2_+(\varphi + \frac{\pi}{2}) \pm \chi^2_-(\varphi + \frac{\pi}{2}) 
\end{array} \right) \nonumber \\
&=&R_2(r) \left(\begin{array}{c}
- \chi^2_-(\varphi) \pm \chi^2_+(\varphi) \\
\chi^2_-(\varphi) \pm \chi^2_+(\varphi) \end{array} \right) = 
\pm \Psi^2_\pm(\vec r), \nonumber \\
^R\Psi^2_\pm(\vec r)&=&R_2(r) \left(\begin{array}{c}
\chi^2_+(- \varphi) \pm \chi^2_-(- \varphi) \\
\chi^2_+(- \varphi) \mp \chi^2_-(- \varphi) \end{array} \right) \nonumber \\
&=&R_2(r) \left(\begin{array}{c}
- \chi^2_+(\varphi) \pm \chi^2_-(\varphi) \\
- \chi^2_+(\varphi) \mp \chi^2_-(\varphi) \end{array} \right) = 
- \Psi^2_\pm(\vec r).
\end{eqnarray}

States with d-wave symmetry can also be constructed for two holes localized on 
a Skyrmion. For example, for $n = 1$, the states 
$\Psi^{\alpha\beta}_{\pm,-1,1,0,2,0}(r_\alpha,\chi_\alpha,r_\beta,\chi_\beta,\gamma)$
and $\Psi^{\alpha\beta}_{\pm,0,2,-1,1,0}(r_\alpha,\chi_\alpha,r_\beta,\chi_\beta,\gamma)$
have d-wave symmetry. Under the symmetries $D_1'$ and $D_2'$ they transform into
$\Psi^{\alpha\beta}_{\mp,-1,1,-2,0,0}(r_\alpha,\chi_\alpha,r_\beta,\chi_\beta,\gamma)$ and
$\Psi^{\alpha\beta}_{\mp,-2,0,-1,1,0}(r_\alpha,\chi_\alpha,r_\beta,\chi_\beta,\gamma)$,
which have s-wave symmetry.  According to eq.(\ref{symtwoholes}), under the 
various symmetries the d-wave states transform as
\begin{eqnarray}
^{D_1'}\Psi^{\alpha\beta}_{\pm,-1,1,0,2,0}(r_\alpha,\chi_\alpha,r_\beta,\chi_\beta,\gamma)
&=&\Psi^{\alpha\beta}_{\mp,-1,1,-2,0,0}(r_\alpha,\chi_\alpha,r_\beta,\chi_\beta,\gamma), 
\nonumber \\
^{D_2'}\Psi^{\alpha\beta}_{\pm,-1,1,0,2,0}(r_\alpha,\chi_\alpha,r_\beta,\chi_\beta,\gamma)
&=&- \Psi^{\alpha\beta}_{\mp,-1,1,-2,0,0}(r_\alpha,\chi_\alpha,r_\beta,\chi_\beta,\gamma),
\nonumber \\
^O\Psi^{\alpha\beta}_{\pm,-1,1,0,2,0}(r_\alpha,\chi_\alpha,r_\beta,\chi_\beta,\gamma)
&=&- \Psi^{\alpha\beta}_{\pm,-1,1,0,2,0}(r_\alpha,\chi_\alpha,r_\beta,\chi_\beta,\gamma), 
\nonumber \\
^R\Psi^{\alpha\beta}_{\pm,-1,1,0,2,0}(r_\alpha,\chi_\alpha,r_\beta,\chi_\beta,\gamma)
&=&\Psi^{\alpha\beta}_{\mp,0,2,-1,1,0}(r_\alpha,\chi_\alpha,r_\beta,\chi_\beta,\gamma),
\nonumber \\
^{D_1'}\Psi^{\alpha\beta}_{\pm,0,2,-1,1,0}(r_\alpha,\chi_\alpha,r_\beta,\chi_\beta,\gamma)
&=&\Psi^{\alpha\beta}_{\mp,-2,0,-1,1,0}(r_\alpha,\chi_\alpha,r_\beta,\chi_\beta,\gamma), 
\nonumber \\
^{D_2'}\Psi^{\alpha\beta}_{\pm,0,2,-1,1,0}(r_\alpha,\chi_\alpha,r_\beta,\chi_\beta,\gamma)
&=&- \Psi^{\alpha\beta}_{\mp,-2,0,-1,1,0}(r_\alpha,\chi_\alpha,r_\beta,\chi_\beta,\gamma),
\nonumber \\
^O\Psi^{\alpha\beta}_{\pm,0,2,-1,1,0}(r_\alpha,\chi_\alpha,r_\beta,\chi_\beta,\gamma)
&=&- \Psi^{\alpha\beta}_{\pm,0,2,-1,1,0}(r_\alpha,\chi_\alpha,r_\beta,\chi_\beta,\gamma), 
\nonumber \\
^R\Psi^{\alpha\beta}_{\pm,0,2,-1,1,0}(r_\alpha,\chi_\alpha,r_\beta,\chi_\beta,\gamma)
&=&\Psi^{\alpha\beta}_{\mp,-1,1,0,2,0}(r_\alpha,\chi_\alpha,r_\beta,\chi_\beta,\gamma).
\end{eqnarray}
Similarly, the s-wave states transform as follows
\begin{eqnarray}
^{D_1'}\Psi^{\alpha\beta}_{\pm,-1,1,-2,0,0}(r_\alpha,\chi_\alpha,r_\beta,\chi_\beta,\gamma)
&=&\Psi^{\alpha\beta}_{\mp,-1,1,0,2,0}(r_\alpha,\chi_\alpha,r_\beta,\chi_\beta,\gamma), 
\nonumber \\
^{D_2'}\Psi^{\alpha\beta}_{\pm,-1,1,-2,0,0}(r_\alpha,\chi_\alpha,r_\beta,\chi_\beta,\gamma)
&=&- \Psi^{\alpha\beta}_{\mp,-1,1,0,2,0}(r_\alpha,\chi_\alpha,r_\beta,\chi_\beta,\gamma),
\nonumber \\
^O\Psi^{\alpha\beta}_{\pm,-1,1,-2,0,0}(r_\alpha,\chi_\alpha,r_\beta,\chi_\beta,\gamma)
&=&\Psi^{\alpha\beta}_{\pm,-1,1,-2,0,0}(r_\alpha,\chi_\alpha,r_\beta,\chi_\beta,\gamma), 
\nonumber \\
^R\Psi^{\alpha\beta}_{\pm,-1,1,-2,0,0}(r_\alpha,\chi_\alpha,r_\beta,\chi_\beta,\gamma)
&=&\Psi^{\alpha\beta}_{\mp,-2,0,-1,1,0}(r_\alpha,\chi_\alpha,r_\beta,\chi_\beta,\gamma),
\nonumber \\
^{D_1'}\Psi^{\alpha\beta}_{\pm,-2,0,-1,1,0}(r_\alpha,\chi_\alpha,r_\beta,\chi_\beta,\gamma)
&=&\Psi^{\alpha\beta}_{\mp,0,2,-1,1,0}(r_\alpha,\chi_\alpha,r_\beta,\chi_\beta,\gamma), 
\nonumber \\
^{D_2'}\Psi^{\alpha\beta}_{\pm,-2,0,-1,1,0}(r_\alpha,\chi_\alpha,r_\beta,\chi_\beta,\gamma)
&=&- \Psi^{\alpha\beta}_{\mp,0,2,-1,1,0}(r_\alpha,\chi_\alpha,r_\beta,\chi_\beta,\gamma),
\nonumber \\
^O\Psi^{\alpha\beta}_{\pm,-2,0,-1,1,0}(r_\alpha,\chi_\alpha,r_\beta,\chi_\beta,\gamma)
&=&\Psi^{\alpha\beta}_{\pm,-2,0,-1,1,0}(r_\alpha,\chi_\alpha,r_\beta,\chi_\beta,\gamma), 
\nonumber \\
^R\Psi^{\alpha\beta}_{\pm,-2,0,-1,1,0}(r_\alpha,\chi_\alpha,r_\beta,\chi_\beta,\gamma)
&=&\Psi^{\alpha\beta}_{\mp,-1,1,-2,0,0}(r_\alpha,\chi_\alpha,r_\beta,\chi_\beta,\gamma).
\end{eqnarray}
Hence, just as for the magnon-mediated excited states, as a consequence of the 
interplay of the various symmetries, s- and d-wave states are again degenerate. 

Alternatively, d-wave states also arise for two holes localized on a Skyrmion 
with winding number $n = 2$. For example, the two states 
$\Psi^{\alpha\beta}_{\pm,-1,2,-1,2,0}(r_\alpha,\chi_\alpha,r_\beta,\chi_\beta,\gamma)$
have d-wave symmetry, and they transform into the states
$\Psi^{\alpha\beta}_{\mp,-2,1,-2,1,0}(r_\alpha,\chi_\alpha,r_\beta,\chi_\beta,\gamma)$,
which again have s-wave symmetry, under $D_1'$ and $D_2'$. According to 
eq.(\ref{symtwoholes}), the d-wave states transform as
\begin{eqnarray}
^{D_1'}\Psi^{\alpha\beta}_{\pm,-1,2,-1,2,0}(r_\alpha,\chi_\alpha,r_\beta,\chi_\beta,\gamma)
&=&\Psi^{\alpha\beta}_{\mp,-2,1,-2,1,0}(r_\alpha,\chi_\alpha,r_\beta,\chi_\beta,\gamma), 
\nonumber \\
^{D_2'}\Psi^{\alpha\beta}_{\pm,-1,2,-1,2,0}(r_\alpha,\chi_\alpha,r_\beta,\chi_\beta,\gamma)
&=&- \Psi^{\alpha\beta}_{\mp,-2,1,-2,1,0}(r_\alpha,\chi_\alpha,r_\beta,\chi_\beta,\gamma),
\nonumber \\
^O\Psi^{\alpha\beta}_{\pm,-1,2,-1,2,0}(r_\alpha,\chi_\alpha,r_\beta,\chi_\beta,\gamma)
&=&- \Psi^{\alpha\beta}_{\pm,-1,2,-1,2,0}(r_\alpha,\chi_\alpha,r_\beta,\chi_\beta,\gamma),
\nonumber \\
^R\Psi^{\alpha\beta}_{\pm,-1,2,-1,2,0}(r_\alpha,\chi_\alpha,r_\beta,\chi_\beta,\gamma)
&=&\Psi^{\alpha\beta}_{\mp,-1,2,-1,2,0}(r_\alpha,\chi_\alpha,r_\beta,\chi_\beta,\gamma),
\end{eqnarray}
while the s-wave states transform as
\begin{eqnarray}
^{D_1'}\Psi^{\alpha\beta}_{\pm,-2,1,-2,1,0}(r_\alpha,\chi_\alpha,r_\beta,\chi_\beta,\gamma)
&=&\Psi^{\alpha\beta}_{\mp,-1,2,-1,2,0}(r_\alpha,\chi_\alpha,r_\beta,\chi_\beta,\gamma), 
\nonumber \\
^{D_2'}\Psi^{\alpha\beta}_{\pm,-2,1,-2,1,0}(r_\alpha,\chi_\alpha,r_\beta,\chi_\beta,\gamma)
&=&- \Psi^{\alpha\beta}_{\mp,-1,2,-1,2,0}(r_\alpha,\chi_\alpha,r_\beta,\chi_\beta,\gamma),
\nonumber \\
^O\Psi^{\alpha\beta}_{\pm,-2,1,-2,1,0}(r_\alpha,\chi_\alpha,r_\beta,\chi_\beta,\gamma)
&=&\Psi^{\alpha\beta}_{\pm,-2,1,-2,1,0}(r_\alpha,\chi_\alpha,r_\beta,\chi_\beta,\gamma), 
\nonumber \\
^R\Psi^{\alpha\beta}_{\pm,-2,1,-2,1,0}(r_\alpha,\chi_\alpha,r_\beta,\chi_\beta,\gamma)
&=&\Psi^{\alpha\beta}_{\mp,-2,1,-2,1,0}(r_\alpha,\chi_\alpha,r_\beta,\chi_\beta,\gamma).
\end{eqnarray}
Depending on the details of the dynamics, whose investigation goes beyond the 
scope of the present paper, it may be possible that the degenerate s- and 
d-wave states have a lower energy than the p-wave states discussed earlier.

\subsection{Possible Implications for Cooper Pair Formation in 
High-Tem\-pe\-ra\-ture Superconductors}

As we have seen, two holes, one of flavor $\alpha$ and one of flavor $\beta$, 
can both get localized in the ground state of a single rotating Skyrmion with
$n = 1$, which turns out to have p-wave symmetry. Alternatively, the holes may
get localized on a rotating $n = 1$ or 2 Skyrmion with s-wave or d-wave 
symmetry. As discussed in Appendix A, two holes of the same flavor can also get
localized on a Skyrmion. It will be the subject of a subsequent publication to 
decide which of the various states is energetically most favorable.

While in this paper we have concentrated on a detailed symmetry analysis, we 
also want to get at least a crude estimate of the binding energy of two-hole
states localized on an $n = 1$ Skyrmion. Ignoring contact interactions between 
the two holes, the total energy of the bound state of two holes and a Skyrmion 
can then be estimated as
\begin{equation}
E_{\text{tot}} = 2 M + 4 \pi \rho_s + 2 E_0,
\end{equation}
while two free holes (not localized on a Skyrmion) just have their rest energy
$2 M$. Using the result of eq.(\ref{bindingE}), the perturbative vacuum thus 
becomes unstable against the formation of two-hole-Skyrmion bound states when
\begin{equation}
4 \pi \rho_s + 2 E_0 < 0 \ \Rightarrow \ 0.270 M' \Lambda^2 > 4 \pi \rho_s.
\end{equation}
Hence, for sufficiently small spin stiffness $\rho_s$, the instability will 
indeed arise. Similar instabilities are related to the formation of spiral
phases in the staggered magnetization order parameter. In particular, in
\cite{Bru07} we have shown that the ground state with a spatially constant 
staggered magnetization becomes unstable against the formation of a 45 degrees 
spiral phase for $M' \Lambda^2 > 4 \pi \rho_s$. Since antiferromagnetism is 
weakened upon doping, $\rho_s$ is expected to eventually go to zero. Before 
this happens, pairs of holes will get localized on a Skyrmion. 

In order to get at least a rough idea of the involved energy scales, let us
estimate the values of the relevant low-energy parameters for realistic
lightly doped quantum antiferromagnets. By comparison with 
\cite{Kuc93,Sus04,Kot05}, where a generalized $t$-$J$ model on a square lattice 
with spacing $a$ was considered at $J/t \approx 0.3$, one obtains the rough 
estimate
\begin{equation}
M' \approx \frac{1}{t a^2} \approx \frac{0.3}{J a^2}, \quad 
\Lambda \approx 2.5 J a.
\end{equation}
It would be interesting and definitely feasible to extract these parameters with
high precision from numerical simulations. In this way, in the Heisenberg model 
(i.e.\ the undoped $t$-$J$ model) very accurate numerical results have been 
obtained for the spin stiffness, the spinwave velocity, and the staggered
magnetization per lattice site \cite{Wie94,San10,Jia11}
\begin{equation}
\rho_s = 0.18081(11) J, \quad c = 1.6586(3) J a, \quad 
{\cal M}_s = 0.30743(1)/a^2.
\end{equation}
Hence, one obtains $0.270 M' \Lambda^2 \approx 0.5 J$ compared to 
$4 \pi \rho_s = 2.2721(1) J$, which implies that two-hole-Skyrmion bound states
are still far from being energetically favorable at zero doping. The exchange 
coupling of undoped $\mbox{La}_2\mbox{CuO}_4$ is $J = 1540(60)$ K \cite{Wie94}. 
A high transition temperature of $T_c \approx 50$ K, and hence
$T_c \approx 0.03 J$, would thus require a two-hole-Skyrmion bound state energy 
of about 
\begin{equation}
4 \pi \rho_s + 2 E_0 = 4 \pi \rho_s - 0.270 M' \Lambda^2 \approx - 0.03 J \ 
\Rightarrow \ \rho_s \approx 0.04 J.
\end{equation}
If doping reduces $\rho_s$ by a factor of about 4 or 5 (and assuming for 
simplicity that the other parameters remain unchanged), the estimated energy 
scales should indeed be of the right magnitude in order to make two-holes 
localized on a rotating Skyrmion a viable candidate for a preformed Cooper pair 
of a high-temperature superconductor. Using eq.(\ref{rho}), one can estimate the
radius of the Skyrmion, which sets the scale for the size of the candidate 
Cooper pair, as $\rho \approx 1/(0.271 M' \Lambda) \approx 5 a$, which again
seems reasonable.

It may involve some wishful thinking to assume that the d-wave state of two
holes localized on an $n = 1$ or 2 Skyrmion will not only turn out to be
energetically favorable, but also ready to condense at sufficiently large 
doping. However, we think that it is worthwhile to take this possibility 
seriously. Deciding whether the radial dynamics favors these states as promising
candidates for a preformed Cooper pair in the pseudo-gap phase is the natural 
next step. The question of condensation is another important issue. 

\section{Conclusions}

We have performed a detailed study of the localization of holes on a Skyrmion 
in an square lattice antiferromagnet. When two holes get localized on the same 
Skyrmion, they form a bound state. Interestingly, in some cases, the quantum 
numbers of these topologically non-trivial bound states are the same as those 
of the topologically trivial bound states resulting from one-magnon exchange 
between two holes. The ground state of two holes weakly bound by one-magnon 
exchange has p-wave symmetry and may evolve into a strongly bound state of two 
holes localized on an $n = 1$ Skyrmion at strong coupling. 

Magnon-mediated two-hole bound states which are excited in the angular motion 
have s- or d-wave symmetry. Remarkably, s- and d-wave states are degenerate due 
to an interplay of the various symmetries. Similarly, there are strongly bound 
states of two holes localized on an $n = 1$ or 2 Skyrmion which also have s- or
d-wave symmetry, and are again degenerate. Which of these states is 
energetically most favorable will be an interesting subject for future studies. 
If a d-wave state turns out to be the ground state at sufficiently strong 
doping, two holes localized on a Skyrmion are a promising candidate for a 
preformed Cooper pair in the pseudo-gap regime. Interestingly, the effective 
theory provides detailed predictions for the anatomy of these objects. In 
particular, their angular structure follows unambiguously from our symmetry 
analysis, and is insensitive to the details of the radial dynamics.

Understanding the dynamical mechanism responsible for high-temperature 
superconductivity has proved to be one of the most challenging problems in 
theoretical physics. While hole pair localization on a rotating Skyrmion may 
ultimately turn out not to be the relevant mechanism, it seems rather promising.
Beyond the symmetry analysis presented here, studying its dynamics in more 
detail is certainly worthwhile.

\section*{Acknowledgments}

We like to thank C.\ Br\"ugger and F.\ K\"ampfer for contributing to the early 
stages of the work reported here. U.-J.\ W.\ likes to thank P.\ A.\ Lee and
F.\ Wilczek for discussions and encouragement at the beginning of the project. 
C.\ P.\ H.\ and N.\ D.\ V.\ thank the Institute for Theoretical Physics at Bern 
University for warm hospitality. C.\ P.\ H.\ gratefully acknowledges financial 
support from the Universidad de Colima. This work is supported by funds provided
by the Schweizerischer Nationalfonds (SNF). In the early phases of this work, 
N.\ D.\ V.\ was supported through an SNF SCOPES grant. The Albert Einstein 
Center for Fundamental Physics at  Bern University is supported by the 
``Innovations- und Kooperationsprojekt C-13'' of the Schweizerische 
Universt\"atskonferenz (SUK/CRUS).

\newpage

\begin{appendix}

\section{Hole Pairs of the Same Flavor}

In this appendix, we consider a pair of holes in a square lattice 
antiferromagnet residing in the same hole pocket. First, we investigate two 
holes localized on a rotating Skyrmion, and then we compare the results with 
the corresponding two-hole magnon-mediated bound states.

\subsection{Schr\"odinger Equation for a Pair of Holes of the Same Flavor
Localized on a Rotating Skyr\-mi\-on}

Let us consider bound states of two holes of the same flavor $f$ localized on 
a rotating Skyrmion. In this case, as a consequence of the Pauli principle, the
holes cannot occupy the same quantum state. We distinguish the holes by an
unphysical label 1 or 2. In order to satisfy the Pauli principle, the wave
function must be anti-symmetric under the exchange of the two labels.

The Hamiltonian for two holes of the same flavor $f$ is then given by 
\begin{equation}
H = H^1 + H^2 + H^\gamma,
\end{equation} 
where
\begin{eqnarray}
H^1&=&\left(\begin{array}{cccc}  
H^f_{++} & 0 & H^f_{+-} & 0 \\ 0 & H^f_{++} & 0 & H^f_{+-} \\
H^f_{-+} & 0 & H^f_{--} & 0 \\ 0 & H^f_{-+} & 0 & H^f_{--} \end{array} \right), 
\quad
H^2 = \left(\begin{array}{cccc}  
H^f_{++} & H^f_{+-} & 0 & 0 \\ H^f_{-+} & H^f_{--} & 0 & 0 \\
0 & 0 & H^f_{++} & H^f_{+-} \\ 0 & 0 & H^f_{-+} & H^f_{--} \end{array} \right), 
\nonumber \\
H^\gamma&=&\left(\begin{array}{cccc}  
H^\gamma_{++++} & 0 & 0 & 0 \\
0 & H^\gamma_{+-+-} & 0 & 0 \\
0 & 0 & H^\gamma_{-+-+} & 0 \\
0 & 0 & 0 & H^\gamma_{----} \end{array} \right),
\end{eqnarray}
with $H^f_{\pm\pm}$ and $H^\gamma_{\pm\pm\pm\pm}$ given in
eq.(\ref{flavorHamiltonians}).

Before anti-symmetrizing the wave function in the artificial labels 1 and 2, we 
ignore the Pauli principle, and make the following ansatz for an energy 
eigenstate of two holes (distinguished by the labels 1 and 2)
\begin{eqnarray}
&&\hskip-1.5cm
\Psi^{ff}_{\sigma,m^1_+,m^1_-,m^2_+,m^2_-,m}(r_1,\chi_1,r_2,\chi_2,\gamma) = \nonumber \\
&&\hskip-1.5cm\left(\begin{array}{c}
\psi_{\sigma,m^1_+,m^1_-,m^2_+,m^2_-,m,++}(r_1,r_2)
\exp\left(i \sigma \left[m^1_+ \chi_1 + m^2_+ \chi_2 - \sigma_f \frac{\pi}{4}
\right]\right) \exp(i \sigma (m - 1) \gamma) \\
\sigma \sigma_f \psi_{\sigma,m^1_+,m^1_-,m^2_+,m^2_-,m,+-}(r_1,r_2)
\exp\left(i \sigma \left[m^1_+ \chi_1 + m^2_- \chi_2\right]\right) 
\exp(i \sigma m \gamma) \\
\sigma \sigma_f \psi_{\sigma,m^1_+,m^1_-,m^2_+,m^2_-,m,-+}(r_1,r_2)
\exp\left(i \sigma \left[m^1_- \chi_1 + m^2_+ \chi_2\right]\right) 
\exp(i \sigma m \gamma) \\
\psi_{\sigma,m^1_+,m^1_-,m^2_+,m^2_-,m,--}(r_1,r_2)
\exp\left(i \sigma \left[m^1_- \chi_1 + m^2_- \chi_2 + \sigma_f \frac{\pi}{4}
\right]\right) \exp(i \sigma (m + 1) \gamma) \end{array}\right). \nonumber \\ \
\end{eqnarray}
As before, this solves the Schr\"odinger equation only if 
$m^i_- - m^i_+ = n + 1$, $i = 1, 2$. In this case, $m$ is again an integer. The
resulting radial Schr\"odinger equation now takes the form
\begin{equation}
\label{twoholeradialsame}
H_r \psi_{\sigma,m^1_+,m^1_-,m^2_+,m^2_-,m}(r_1,r_2) =
E_{\sigma,m^1_+,m^1_-,m^2_+,m^2_-,m} \psi_{\sigma,m^1_+,m^1_-,m^2_+,m^2_-,m}(r_1,r_2),
\end{equation}
with
\begin{equation}
\psi_{\sigma,m^1_+,m^1_-,m^2_+,m^2_-,m}(r_1,r_2) =
\left(\begin{array}{c}
\psi_{\sigma,m^1_+,m^1_-,m^2_+,m^2_-,m,++}(r_1,r_2) \\
\psi_{\sigma,m^1_+,m^1_-,m^2_+,m^2_-,m,+-}(r_1,r_2) \\
\psi_{\sigma,m^1_+,m^1_-,m^2_+,m^2_-,m,-+}(r_1,r_2) \\
\psi_{\sigma,m^1_+,m^1_-,m^2_+,m^2_-,m,--}(r_1,r_2) \end{array}\right).
\end{equation}
The radial Hamiltonian is given by
\begin{equation}
H_r = H_r^1 + H_r^2 + H_r^\gamma,
\end{equation} 
with
\begin{eqnarray}
H_r^1&=&\left(\begin{array}{cccc}  
H^1_{r++} & 0 & H^1_{r+-} & 0 \\ 0 & H^1_{r++} & 0 & H^1_{r+-} \\
H^1_{r-+} & 0 & H^1_{r--} & 0 \\ 0 & H^1_{r-+} & 0 & H^1_{r--} \end{array} \right), 
\nonumber \\
H_r^2&=&\left(\begin{array}{cccc}  
H^2_{r++} & H^2_{r+-} & 0 & 0 \\ H^2_{r-+} & H^2_{r--} & 0 & 0 \\
0 & 0 & H^2_{r++} & H^2_{r+-} \\ 0 & 0 & H^2_{r-+} & H^2_{r--} \end{array} \right), 
\nonumber \\
H_r^\gamma&=&\left(\begin{array}{cccc}  
H^\gamma_{r++++} & 0 & 0 & 0 \\
0 & H^\gamma_{r+-+-} & 0 & 0 \\
0 & 0 & H^\gamma_{r-+-+} & 0 \\
0 & 0 & 0 & H^\gamma_{r----} \end{array} \right).
\end{eqnarray}
The matrix elements of the fermionic part of the radial Hamiltonian are given by
\begin{eqnarray}
H^i_{r++}&=&- \frac{1}{2 M'} 
\left[\p_{r_i}^2 + \frac{1}{r_i} \p_{r_i} - \frac{1}{r_i^2} 
\left(m^i_+ + \frac{n \rho^{2n}}{r_i^{2n} + \rho^{2n}}\right)^2\right], 
\nonumber \\
H^i_{r+-}&=&H^i_{r-+} = 
\sqrt{2} \Lambda \frac{n r_i^{n-1} \rho^n}{r_i^{2n} + \rho^{2n}}, \nonumber \\
H^i_{r--}&=&- \frac{1}{2 M'} 
\left[\p_{r_i}^2 + \frac{1}{r_i} \p_{r_i} - \frac{1}{r_i^2} 
\left(m^i_- - \frac{n \rho^{2n}}{r_i^{2n} + \rho^{2n}}\right)^2\right],
\end{eqnarray}
while the rotational Skyrmion contributions are given by
\begin{eqnarray}
H^\gamma_{r++++}&=&\frac{n^2}{2 {\cal D}(\rho) \rho^2} 
\left(m + \sigma n \frac{\Theta}{2 \pi} - 1 -  
\frac{\rho^{2n}}{r_1^{2n} + \rho^{2n}} - 
\frac{\rho^{2n}}{r_2^{2n} + \rho^{2n}}\right)^2, \nonumber \\
H^\gamma_{r+-+-}&=&\frac{n^2}{2 {\cal D}(\rho) \rho^2} 
\left(m + \sigma n \frac{\Theta}{2 \pi} -  
\frac{\rho^{2n}}{r_1^{2n} + \rho^{2n}} +
\frac{\rho^{2n}}{r_2^{2n} + \rho^{2n}}\right)^2, \nonumber \\
H^\gamma_{r-+-+}&=&\frac{n^2}{2 {\cal D}(\rho) \rho^2} 
\left(m + \sigma n \frac{\Theta}{2 \pi} +  
\frac{\rho^{2n}}{r_1^{2n} + \rho^{2n}} -
\frac{\rho^{2n}}{r_2^{2n} + \rho^{2n}}\right)^2, \nonumber \\
H^\gamma_{r----}&=&\frac{n^2}{2 {\cal D}(\rho) \rho^2} 
\left(m + \sigma n \frac{\Theta}{2 \pi} + 1 +
\frac{\rho^{2n}}{r_1^{2n} + \rho^{2n}} +
\frac{\rho^{2n}}{r_2^{2n} + \rho^{2n}}\right)^2.
\end{eqnarray}

\subsection{Symmetry Properties of a Pair of Holes with the Same Flavor 
Localized on a Skyrmion}

The spin operator $I$ is again given by eq.(\ref{spinoperator}), such that 
\begin{eqnarray}
&&I \Psi^{ff}_{\sigma,m^1_+,m^1_-,m^2_+,m^2_-,m}(r_1,\chi_1,r_2,\chi_2,\gamma) = 
\nonumber \\
&&\left(m + \sigma n \frac{\Theta}{2 \pi}\right) 
\Psi^{ff}_{\sigma,m^1_+,m^1_-,m^2_+,m^2_-,m}(r_1,\chi_1,r_2,\chi_2,\gamma).
\end{eqnarray}
Since $m$ is an integer, at least for $\Theta = 0$, the state with two holes of
the same flavor localized on a Skyrmion again has integer spin.

The symmetries $D_i'$, $O$, and $R$ act on the two-hole wave function
\begin{equation}
\Psi^{ff}_{\sigma,n}(r_1,\chi_1,r_2,\chi_2,\gamma) =
\left(\begin{array}{c} 
\Psi^{ff}_{\sigma,n,++}(r_1,\chi_1,r_2,\chi_2,\gamma) \\
\Psi^{ff}_{\sigma,n,+-}(r_1,\chi_1,r_2,\chi_2,\gamma) \\
\Psi^{ff}_{\sigma,n,-+}(r_1,\chi_1,r_2,\chi_2,\gamma) \\
\Psi^{ff}_{\sigma,n,--}(r_1,\chi_1,r_2,\chi_2,\gamma) \end{array}\right)
\end{equation}
as follows
\begin{eqnarray}
&&^{D_i'}\Psi^{ff}_{\sigma,n}(r_1,\chi_1,r_2,\chi_2,\gamma) =
\exp(2 i k_i^f a) \left(\begin{array}{c} 
\Psi^{ff}_{\sigma,n,--}(r_1,\chi_1,r_2,\chi_2,\gamma) \\
- \Psi^{ff}_{\sigma,n,-+}(r_1,\chi_1,r_2,\chi_2,\gamma) \\
- \Psi^{ff}_{\sigma,n,+-}(r_1,\chi_1,r_2,\chi_2,\gamma) \\
\Psi^{ff}_{\sigma,n,++}(r_1,\chi_1,r_2,\chi_2,\gamma)
\end{array}\right), \nonumber \\
&&^O\Psi^{ff}_{\sigma,n}(r_1,\chi_1,r_2,\chi_2,\gamma) =
\left(\begin{array}{c} 
\Psi^{ff}_{\sigma,n,++}(r_1,\chi_1 + \frac{\pi}{2},
r_2,\chi_2 + \frac{\pi}{2},\gamma - n \frac{\pi}{2}) \\
\sigma_f \Psi^{ff}_{\sigma,n,+-}(r_1,\chi_1 + \frac{\pi}{2},
r_2,\chi_2 + \frac{\pi}{2},\gamma - n \frac{\pi}{2}) \\
\sigma_f \Psi^{ff}_{\sigma,n,-+}(r_1,\chi_1 + \frac{\pi}{2},
r_2,\chi_2 + \frac{\pi}{2},\gamma - n \frac{\pi}{2}) \\
\Psi^{ff}_{\sigma,n,--}(r_1,\chi_1 + \frac{\pi}{2},
r_2,\chi_2 + \frac{\pi}{2},\gamma - n \frac{\pi}{2}) \end{array}\right), 
\nonumber \\
&&^R\Psi^{ff}_{\sigma,n}(r_1,\chi_1,r_2,\chi_2,\gamma) =
\left(\begin{array}{c} 
\Psi^{ff}_{\sigma,n,++}(r_1,-\chi_1,r_2,-\chi_2,-\gamma) \\
\Psi^{ff}_{\sigma,n,+-}(r_1,-\chi_1,r_2,-\chi_2,-\gamma) \\
\Psi^{ff}_{\sigma,n,-+}(r_1,-\chi_1,r_2,-\chi_2,-\gamma) \\
\Psi^{ff}_{\sigma,n,--}(r_1,-\chi_1,r_2,-\chi_2,-\gamma)
\end{array}\right).
\end{eqnarray}

For the two-hole energy eigenstates this implies 
\begin{eqnarray}
\label{symtwoholessame}
^{D_i'}\Psi^{ff}_{\sigma,m^1_+,m^1_-,m^2_+,m^2_-,m}
(r_1,\chi_1,r_2,\chi_2,\gamma)\!\!\!\!&=&\!\!\!\!
- \Psi^{ff}_{-\sigma,-m^1_-,-m^1_+,-m^2_-,-m^2_+,-m}
(r_1,\chi_1,r_2,\chi_2,\gamma), \nonumber \\
^O\Psi^{\alpha\alpha}_{\sigma,m^1_+,m^1_-,m^2_+,m^2_-,m}
(r_1,\chi_1,r_2,\chi_2,\gamma)\!\!\!\!&=&\!\!\!\!
- \exp\left(i \sigma [m^1_+ + m^2_- - m n] \frac{\pi}{2}\right)
\nonumber \\
&\times&\!\!\!\!\Psi^{\beta\beta}_{\sigma,m^1_+,m^1_-,m^2_+,m^2_-,m}
(r_1,\chi_1,r_2,\chi_2,\gamma), \nonumber \\
^O\Psi^{\beta\beta}_{\sigma,m^1_+,m^1_-,m^2_+,m^2_-,m}
(r_1,\chi_1,r_2,\chi_2,\gamma)\!\!\!\!&=&\!\!\!\!
\exp\left(i \sigma [m^1_+ + m^2_- - m n] \frac{\pi}{2}\right)
\nonumber \\
&\times&\!\!\!\!\Psi^{\alpha\alpha}_{\sigma,m^1_+,m^1_-,m^2_+,m^2_-,m}
(r_1,\chi_1,r_2,\chi_2,\gamma), \nonumber \\
^R\Psi^{\alpha\alpha}_{\sigma,m^1_+,m^1_-,m^2_+,m^2_-,m}
(r_1,\chi_1,r_2,\chi_2,\gamma)\!\!\!\!&=&\!\!\!\!
\Psi^{\beta\beta}_{-\sigma,m^1_+,m^1_-,m^2_+,m^2_-,m}
(r_1,\chi_1,r_2,\chi_2,\gamma), \nonumber \\
^R\Psi^{\beta\beta}_{\sigma,m^1_+,m^1_-,m^2_+,m^2_-,m}
(r_1,\chi_1,r_2,\chi_2,\gamma)\!\!\!\!&=&\!\!\!\!
\Psi^{\alpha\alpha}_{-\sigma,m^1_+,m^1_-,m^2_+,m^2_-,m}
(r_1,\chi_1,r_2,\chi_2,\gamma).
\end{eqnarray}
Here we have again assumed an appropriate phase convention for the radial wave 
function $\psi_{\sigma,m^1_+,m^1_-,m^2_+,m^2_-,m}(r_1,r_2)$. In the context of the shift
symmetries $D_i'$ we have used
\begin{eqnarray}
&&\psi_{\sigma,m^1_+,m^1_-,m^2_+,m^2_-,m,--}(r_1,r_2) = 
\psi_{-\sigma,-m^1_-,-m^1_+,-m^2_-,-m^2_+,-m,++}(r_1,r_2), \nonumber \\
&&\psi_{\sigma,m^1_+,m^1_-,m^2_+,m^2_-,m,-+}(r_1,r_2) = 
\psi_{-\sigma,-m^1_-,-m^1_+,-m^2_-,-m^2_+,-m,+-}(r_1,r_2), \nonumber \\
&&\psi_{\sigma,m^1_+,m^1_-,m^2_+,m^2_-,m,+-}(r_1,r_2) = 
\psi_{-\sigma,-m^1_-,-m^1_+,-m^2_-,-m^2_+,-m,-+}(r_1,r_2), \nonumber \\
&&\psi_{\sigma,m^1_+,m^1_-,m^2_+,m^2_-,m,++}(r_1,r_2) = 
\psi_{-\sigma,-m^1_-,-m^1_+,-m^2_-,-m^2_+,-m,--}(r_1,r_2).
\end{eqnarray}
These relations follow from the symmetries of the radial Schr\"odinger equation
(\ref{twoholeradialsame}). In the context of the reflection symmetry $R$ we 
have used
\begin{eqnarray}
\label{Rsymsame}
&&\psi_{\sigma,m^1_+,m^1_-,m^2_+,m^2_-,m,++}(r_1,r_2) = 
\psi_{-\sigma,m^1_+,m^1_-,m^2_+,m^2_-,m,++}(r_1,r_2), \nonumber \\
&&\psi_{\sigma,m^1_+,m^1_-,m^2_+,m^2_-,m,+-}(r_1,r_2) = 
\psi_{-\sigma,m^1_+,m^1_-,m^2_+,m^2_-,m,+-}(r_1,r_2), \nonumber \\
&&\psi_{\sigma,m^1_+,m^1_-,m^2_+,m^2_-,m,-+}(r_1,r_2) = 
\psi_{-\sigma,m^1_+,m^1_-,m^2_+,m^2_-,m,-+}(r_1,r_2), \nonumber \\
&&\psi_{\sigma,m^1_+,m^1_-,m^2_+,m^2_-,m,--}(r_1,r_2) = 
\psi_{-\sigma,m^1_+,m^1_-,m^2_+,m^2_-,m,--}(r_1,r_2).
\end{eqnarray}
The relations in eq.(\ref{Rsymsame}) follow from the symmetries of the radial 
Schr\"odinger equation (\ref{twoholeradialsame}) for $\Theta = 0$. As before,
for $\Theta \neq 0$ or $\pi$, the Hopf term explicitly breaks the reflection 
symmetry.

Let us now impose the Pauli principle by explicitly anti-symmetrizing the wave
function in the artificial indices 1 and 2. For this purpose we act with the
pair permutation $P$, i.e.
\begin{equation}
^P\Psi^{ff}_{\sigma,n}(r_1,\chi_1,r_2,\chi_2,\gamma) =
\left(\begin{array}{c} 
\Psi^{ff}_{\sigma,n,++}(r_2,\chi_2,r_1,\chi_1,\gamma) \\
\Psi^{ff}_{\sigma,n,-+}(r_2,\chi_2,r_1,\chi_1,\gamma) \\
\Psi^{ff}_{\sigma,n,+-}(r_2,\chi_2,r_1,\chi_1,\gamma) \\
\Psi^{ff}_{\sigma,n,--}(r_2,\chi_2,r_1,\chi_1,\gamma) \end{array}\right).
\end{equation}
For an energy eigenstate this implies
\begin{equation}
^P\Psi^{ff}_{\sigma,m^1_+,m^1_-,m^2_+,m^2_-,m}(r_1,\chi_1,r_2,\chi_2,\gamma) =
\Psi^{ff}_{\sigma,m^2_+,m^2_-,m^1_+,m^1_-,m}(r_1,\chi_1,r_2,\chi_2,\gamma).
\end{equation}
Here we have assumed a symmetric radial wave function, i.e.
\begin{eqnarray}
&&\psi_{\sigma,m^1_+,m^1_-,m^2_+,m^2_-,m,++}(r_2,r_1) = 
\psi_{\sigma,m^2_+,m^2_-,m^1_+,m^1_-,m,++}(r_1,r_2), \nonumber \\
&&\psi_{\sigma,m^1_+,m^1_-,m^2_+,m^2_-,m,-+}(r_2,r_1) = 
\psi_{\sigma,m^2_+,m^2_-,m^1_+,m^1_-,m,+-}(r_1,r_2), \nonumber \\
&&\psi_{\sigma,m^1_+,m^1_-,m^2_+,m^2_-,m,+-}(r_2,r_1) = 
\psi_{\sigma,m^2_+,m^2_-,m^1_+,m^1_-,m,-+}(r_1,r_2), \nonumber \\
&&\psi_{\sigma,m^1_+,m^1_-,m^2_+,m^2_-,m,--}(r_2,r_1) = 
\psi_{\sigma,m^2_+,m^2_-,m^1_+,m^1_-,m,--}(r_1,r_2).
\end{eqnarray}
The properly anti-symmetrized wave function now takes the form
\begin{equation}
\widetilde\Psi^{ff}_{\sigma,n}(r_1,\chi_1,r_2,\chi_2,\gamma) = \frac{1}{\sqrt{2}}
\left[\Psi^{ff}_{\sigma,n}(r_1,\chi_1,r_2,\chi_2,\gamma) -
^P\Psi^{ff}_{\sigma,n}(r_1,\chi_1,r_2,\chi_2,\gamma)\right].
\end{equation}
For an energy eigenstate this implies
\begin{eqnarray}
&&\hspace{-2cm}
\widetilde\Psi^{ff}_{\sigma,m^1_+,m^1_-,m^2_+,m^2_-,m}(r_1,\chi_1,r_2,\chi_2,\gamma) = 
\nonumber \\
&&\hspace{-2cm} \frac{1}{\sqrt{2}}
\left[\Psi^{ff}_{\sigma,m^1_+,m^1_-,m^2_+,m^2_-,m}(r_1,\chi_1,r_2,\chi_2,\gamma) -
\Psi^{ff}_{\sigma,m^2_+,m^2_-,m^1_+,m^1_-,m}(r_1,\chi_1,r_2,\chi_2,\gamma)\right].
\end{eqnarray}
As expected, in order to obtain a non-vanishing wave function, the two sets of 
quantum numbers $m^1_+, m^1_-$ and $m^2_+, m^2_-$ must be different, because
otherwise two identical fermions would occupy the same single particle state. If
one would consider an anti-symmetric radial wave function, one could allow
$m^1_+ = m^2_+$ and $m^1_- = m^2_-$. 

Based on eq.(\ref{symtwoholessame}), the properly anti-symmetrized two-hole 
energy eigenstates transform as follows
\begin{eqnarray}
\label{symtwoholessameanti}
^{D_i'}\widetilde\Psi^{ff}_{\sigma,m^1_+,m^1_-,m^2_+,m^2_-,m}
(r_1,\chi_1,r_2,\chi_2,\gamma)\!\!\!\!&=&\!\!\!\!
- \widetilde\Psi^{ff}_{-\sigma,-m^1_-,-m^1_+,-m^2_-,-m^2_+,-m}
(r_1,\chi_1,r_2,\chi_2,\gamma), \nonumber \\
^O\widetilde\Psi^{\alpha\alpha}_{\sigma,m^1_+,m^1_-,m^2_+,m^2_-,m}
(r_1,\chi_1,r_2,\chi_2,\gamma)\!\!\!\!&=&\!\!\!\!
- \exp\left(i \sigma [m^1_+ + m^2_- - m n] \frac{\pi}{2}\right)
\nonumber \\
&\times&\!\!\!\!\widetilde\Psi^{\beta\beta}_{\sigma,m^1_+,m^1_-,m^2_+,m^2_-,m}
(r_1,\chi_1,r_2,\chi_2,\gamma), \nonumber \\
^O\widetilde\Psi^{\beta\beta}_{\sigma,m^1_+,m^1_-,m^2_+,m^2_-,m}
(r_1,\chi_1,r_2,\chi_2,\gamma)\!\!\!\!&=&\!\!\!\!
\exp\left(i \sigma [m^1_+ + m^2_- - m n] \frac{\pi}{2}\right)
\nonumber \\
&\times&\!\!\!\!\widetilde\Psi^{\alpha\alpha}_{\sigma,m^1_+,m^1_-,m^2_+,m^2_-,m}
(r_1,\chi_1,r_2,\chi_2,\gamma), \nonumber \\
^R\widetilde\Psi^{\alpha\alpha}_{\sigma,m^1_+,m^1_-,m^2_+,m^2_-,m}
(r_1,\chi_1,r_2,\chi_2,\gamma)\!\!\!\!&=&\!\!\!\!
\widetilde\Psi^{\beta\beta}_{-\sigma,m^1_+,m^1_-,m^2_+,m^2_-,m}
(r_1,\chi_1,r_2,\chi_2,\gamma), \nonumber \\
^R\widetilde\Psi^{\beta\beta}_{\sigma,m^1_+,m^1_-,m^2_+,m^2_-,m}
(r_1,\chi_1,r_2,\chi_2,\gamma)\!\!\!\!&=&\!\!\!\!
\widetilde\Psi^{\alpha\alpha}_{-\sigma,m^1_+,m^1_-,m^2_+,m^2_-,m}
(r_1,\chi_1,r_2,\chi_2,\gamma).
\end{eqnarray}
In order to show this for the rotation $O$, we have used 
$m^1_+ + m^2_- = m^2_+ + m^1_-$. 

Finally, let us combine states with flavors $\alpha\alpha$ and $\beta\beta$ to 
eigenstates of $O$,
\begin{eqnarray}
&&\hspace{-2cm}
\widetilde\Psi^\pm_{\sigma,m^1_+,m^1_-,m^2_+,m^2_-,m}(r_1,\chi_1,r_2,\chi_2,\gamma) = 
\nonumber \\
&&\hspace{-2cm}
\frac{1}{\sqrt{2}} \left[\widetilde\Psi^{\alpha\alpha}_{\sigma,m^1_+,m^1_-,m^2_+,m^2_-,m}
(r_1,\chi_1,r_2,\chi_2,\gamma) \pm i 
\widetilde\Psi^{\beta\beta}_{\sigma,m^1_+,m^1_-,m^2_+,m^2_-,m}
(r_1,\chi_1,r_2,\chi_2,\gamma)\right],
\end{eqnarray}
which transform as
\begin{eqnarray}
\label{symtwoholessamerot}
^{D_i'}\widetilde\Psi^\pm_{\sigma,m^1_+,m^1_-,m^2_+,m^2_-,m}
(r_1,\chi_1,r_2,\chi_2,\gamma)\!\!\!\!&=&\!\!\!\!
- \widetilde\Psi^\pm_{-\sigma,-m^1_-,-m^1_+,-m^2_-,-m^2_+,-m}
(r_1,\chi_1,r_2,\chi_2,\gamma), \nonumber \\
^O\widetilde\Psi^\pm_{\sigma,m^1_+,m^1_-,m^2_+,m^2_-,m}
(r_1,\chi_1,r_2,\chi_2,\gamma)\!\!\!\!&=&\!\!\!\!
\pm i \exp\left(i \sigma [m^1_+ + m^2_- - m n] \frac{\pi}{2}\right)
\nonumber \\
&\times&\!\!\!\!\widetilde\Psi^\pm_{\sigma,m^1_+,m^1_-,m^2_+,m^2_-,m}
(r_1,\chi_1,r_2,\chi_2,\gamma), \nonumber \\
^R\widetilde\Psi^\pm_{\sigma,m^1_+,m^1_-,m^2_+,m^2_-,m}
(r_1,\chi_1,r_2,\chi_2,\gamma)\!\!\!\!&=&\!\!\!\!
\pm i \widetilde\Psi^\mp_{-\sigma,m^1_+,m^1_-,m^2_+,m^2_-,m}
(r_1,\chi_1,r_2,\chi_2,\gamma).
\end{eqnarray}

The lowest energy states in the same flavor channel are expected to correspond 
to $m^1_+ = -1, m^1_- = 1, m^2_+ = -2, m^2_- = 0, m = 0$ or
$m^1_+ = -1, m^1_- = 1, m^2_+ = 0, m^2_- = 2, m = 0$. These states transform as
\begin{eqnarray}
^{D_i'}\widetilde\Psi^\pm_{\sigma,-1,1,-2,0,0}
(r_1,\chi_1,r_2,\chi_2,\gamma)\!\!\!\!&=&\!\!\!\!
- \widetilde\Psi^\pm_{-\sigma,-1,1,0,2,0}(r_1,\chi_1,r_2,\chi_2,\gamma), 
\nonumber \\
^O\widetilde\Psi^\pm_{\sigma,-1,1,-2,0,0}
(r_1,\chi_1,r_2,\chi_2,\gamma)\!\!\!\!&=&\!\!\!\!
\pm \sigma \widetilde\Psi^\pm_{\sigma,-1,1,-2,0,0}
(r_1,\chi_1,r_2,\chi_2,\gamma), 
\nonumber \\
^R\widetilde\Psi^\pm_{\sigma,-1,1,-2,0,0}
(r_1,\chi_1,r_2,\chi_2,\gamma)\!\!\!\!&=&\!\!\!\!
\pm i \widetilde\Psi^\mp_{-\sigma,-1,1,-2,0,0}(r_1,\chi_1,r_2,\chi_2,\gamma),
\nonumber \\
^{D_i'}\widetilde\Psi^\pm_{\sigma,-1,1,0,2,0}
(r_1,\chi_1,r_2,\chi_2,\gamma)\!\!\!\!&=&\!\!\!\!
- \widetilde\Psi^\pm_{-\sigma,-1,1,-2,0,0}(r_1,\chi_1,r_2,\chi_2,\gamma), 
\nonumber \\
^O\widetilde\Psi^\pm_{\sigma,-1,1,0,2,0}
(r_1,\chi_1,r_2,\chi_2,\gamma)\!\!\!\!&=&\!\!\!\!
\mp \sigma \widetilde\Psi^\pm_{\sigma,-1,1,0,2,0}
(r_1,\chi_1,r_2,\chi_2,\gamma), 
\nonumber \\
^R\widetilde\Psi^\pm_{\sigma,-1,1,0,2,0}
(r_1,\chi_1,r_2,\chi_2,\gamma)\!\!\!\!&=&\!\!\!\!
\pm i \widetilde\Psi^\mp_{-\sigma,-1,1,0,2,0}(r_1,\chi_1,r_2,\chi_2,\gamma).
\end{eqnarray}
This implies that the states
$\widetilde\Psi^+_{+,-1,1,-2,0,0}(r_1,\chi_1,r_2,\chi_2,\gamma)$, 
$\widetilde\Psi^-_{-,-1,1,-2,0,0}(r_1,\chi_1,r_2,\chi_2,\gamma)$, 
$\widetilde\Psi^+_{-,-1,1,0,2,0}(r_1,\chi_1,r_2,\chi_2,\gamma)$, 
$\widetilde\Psi^-_{+,-1,1,0,2,0}(r_1,\chi_1,r_2,\chi_2,\gamma)$ are s-waves, while 
the states
$\widetilde\Psi^+_{+,-1,1,0,2,0}(r_1,\chi_1,r_2,\chi_2,\gamma)$, 
$\widetilde\Psi^-_{-,-1,1,0,2,0}(r_1,\chi_1,r_2,\chi_2,\gamma)$, 
$\widetilde\Psi^+_{-,-1,1,-2,0,0}(r_1,\chi_1,r_2,\chi_2,\gamma)$, 
$\widetilde\Psi^-_{+,-1,1,-2,0,0}(r_1,\chi_1,r_2,\chi_2,\gamma)$ are d-waves.

\subsection{Comparison with Magnon-Mediated Two-Hole Bound States of the Same 
Flavor}

In \cite{Bru06} states of two holes of the same flavor bound by one-magnon 
exchange have also been investigated. Here we summarize as well as extend some 
of the relevant results.
\begin{figure}[tb]
\begin{center}
\vspace{-0.3cm}
\epsfig{file=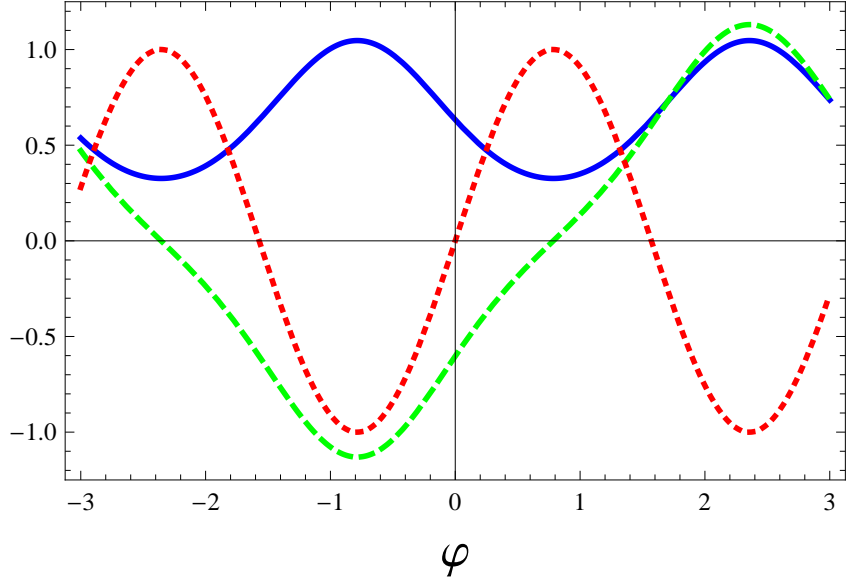,width=11cm}
\end{center}
\caption{\it Angular wave functions 
$\mbox{ce}_0(\varphi - \frac{\pi}{4},\frac{1}{2} M' \gamma)$ (solid curve) and 
$\mbox{se}_1(\varphi - \frac{\pi}{4},\frac{1}{2} M' \gamma)$ (dashed curve) as
well as angle-dependence $\sin(2 \varphi)$ of the potential (dotted curve) for 
two holes of flavor $\alpha$ residing in a circular hole pocket
($M' \Lambda^2/4 \pi \rho_s = 1.25$).}
\end{figure}
We consider two holes of the same flavor $f$ with opposite spins $+$ and $-$. 
In the rest frame the wave function depends on the distance vector $\vec r$ 
which points from the spin $+$ hole to the spin $-$ hole. Since magnon exchange 
is accompanied by a spin-flip, the vector $\vec r$ changes its direction in the 
magnon exchange process. The Schr\"odinger equation thus takes the form
\begin{equation}
- \frac{1}{M'} \Delta \Psi(\vec r) + V^{ff}(\vec r) \Psi(- \vec r) =
E \Psi(\vec r).
\end{equation}
The one-magnon exchange potential for two holes of the same flavor is given by
\begin{equation}
V^{\alpha\alpha}(\vec r) = 
\frac{\Lambda^2}{2 \pi \rho_s} \frac{\sin(2 \varphi)}{r^2}, \quad
V^{\beta\beta}(\vec r) = 
- \frac{\Lambda^2}{2 \pi \rho_s} \frac{\sin(2 \varphi)}{r^2}.
\end{equation}
We make a separation ansatz
\begin{equation}
\Psi(\vec r) = R'(r) \chi'(\varphi).
\end{equation}
The ground state is even with respect to the reflection of $\vec r$ to 
$- \vec r$, i.e.\
\begin{equation}
\chi'^1(\varphi + \pi) = \chi'^1(\varphi).
\end{equation}
The angular part of the Schr\"odinger equation then takes the form
\begin{equation}
\label{Mathieu}
- \frac{d^2\chi'^1_\pm(\varphi)}{d\varphi^2} \pm
\frac{M' \Lambda^2}{2 \pi \rho_s} \sin(2 \varphi) \chi'^1_\pm(\varphi) = 
- \lambda_1 \chi'^1_\pm(\varphi).
\end{equation}
Here $+$ and $-$ are associated with an $\alpha\alpha$ and a $\beta\beta$ pair,
respectively. Again, eq.(\ref{Mathieu}) is a Mathieu equation. The ground state 
with eigenvalue $- \lambda_1$ takes the form
\begin{eqnarray}
&&\chi'^1_\pm(\varphi) = \chi^1_\pm(\varphi - \frac{\pi}{4}) = 
\frac{1}{\sqrt{\pi}} \mbox{ce}_0 \big(\varphi - \frac{\pi}{4},
\pm \frac{M' \Lambda^2}{4 \pi \rho_s}\big), \nonumber \\
&&\lambda_1 = \frac{1}{2} \left(\frac{M' \Lambda^2}{4 \pi \rho_s}\right)^2 +
{\cal O}(\Lambda^8).
\end{eqnarray}

The first excited states are odd with respect to the reflection of $\vec r$ to 
$- \vec r$, i.e.\
\begin{equation}
\chi'^2_\pm(\varphi + \pi) = - \chi'^2_\pm(\varphi),
\end{equation}
and the angular part of the Schr\"odinger equation now reads
\begin{equation}
- \frac{d^2\chi'^2_\mp(\varphi)}{d\varphi^2} \mp
\frac{M' \Lambda^2}{2 \pi \rho_s} \sin(2 \varphi) \chi'^2_\mp(\varphi) = 
- \lambda_2 \chi'^2_\mp(\varphi).
\end{equation}
Now $-$ and $+$ are associated with an $\alpha\alpha$ and a $\beta\beta$ pair,
respectively. The excited states with eigenvalue $- \lambda_2$ are given by
\begin{eqnarray}
&&\chi'^2_+(\varphi) = \chi^2_+(\varphi - \frac{\pi}{4}) =
\frac{1}{\sqrt{\pi}} \mbox{se}_1 \big(\varphi - \frac{\pi}{4},
\frac{M' \Lambda^2}{4 \pi \rho_s}\big), \nonumber \\
&&\chi'^2_-(\varphi) = \chi^2_-(\varphi - \frac{\pi}{4}) =
- \frac{1}{\sqrt{\pi}} \mbox{ce}_1 \big(\varphi - \frac{\pi}{4},
- \frac{M' \Lambda^2}{4 \pi \rho_s}\big), \nonumber \\
&&\lambda_2 = - 1 + \frac{M' \Lambda^2}{4 \pi \rho_s} +
\frac{1}{8} \left(\frac{M' \Lambda^2}{4 \pi \rho_s}\right)^2 -
\frac{1}{64} \left(\frac{M' \Lambda^2}{4 \pi \rho_s}\right)^3 +
{\cal O}(\Lambda^8).
\end{eqnarray}
The angular wave functions for the ground state and for the first excited state 
together with the angular dependence of the one-magnon exchange potential are 
shown in figure 5.

As before, the radial Schr\"odinger equation takes the form of 
eq.(\ref{radial}). Again, the short-distance repulsion between two holes is
modeled by a hard core of radius $r_0'$, i.e.\ $R'(r_0') = 0$. The value of 
$r_0'$ may, however, differ from $r_0$ in the $\alpha \beta$ case. The radial
wave functions are thus given by
\begin{equation}
R'_i(r) = A'_i K_\nu \big( \sqrt{M' |E'_{ik}|} r \big), \quad k = 1,2,3,\dots, 
\quad \nu = i \sqrt{\lambda_i},
\end{equation}
and the energy is determined from 
$K_\nu \big(\sqrt{M' |E'_{ik}|} r'_0 \big) = 0$.

There are two degenerate states --- one for an $\alpha \alpha$ and one for a 
$\beta \beta$ pair, which are eigenstates of flavor related to each other by a 
90 degrees rotation. The two degenerate states can be combined to eigenstates 
of the rotation symmetry $O$. For this purpose, we construct the 2-component
wave functions
\begin{equation}
\Psi'^1_\pm(\vec r) = R'_1(r) \left(\begin{array}{c} 
\chi'^1_+(\varphi) \\ \pm i \chi'^1_-(\varphi) \end{array}\right), \quad
\Psi'^2_\pm(\vec r) = R'_2(r) \left(\begin{array}{c} 
\chi'^2_-(\varphi) \\ \pm \chi'^2_+(\varphi) \end{array}\right),
\end{equation}
whose first component represents the $\alpha\alpha$ and whose second component
represents the $\beta\beta$ pair. Under the various symmetries, the two
degenerate ground states transform as
\begin{eqnarray}
^{D_i'}\Psi'^1_\pm(\vec r)&=&R'_1(r) \left(\begin{array}{c}
\chi'^1_+(\varphi + \pi) \\ \pm i \chi'^1_-(\varphi + \pi) \end{array} \right) =
R'_1(r) \left(\begin{array}{c}
\chi'^1_+(\varphi) \\ \pm i \chi'^1_-(\varphi) \end{array} \right) = 
\Psi'^1_\pm(\vec r), \nonumber \\
^O\Psi'^1_\pm(\vec r)&=&R'_1(r) \left(\begin{array}{c}
\pm i \chi'^1_-(\varphi + \frac{\pi}{2}) \\ - \chi'^1_+(\varphi + \frac{\pi}{2})
\end{array} \right) = R'_1(r) \left(\begin{array}{c}
\pm i \chi'^1_+(\varphi) \\ - \chi'^1_-(\varphi) \end{array} \right) = 
\pm i \Psi'^1_\pm(\vec r), \nonumber \\
^R\Psi'^1_\pm(\vec r)&=&R'_1(r) \left(\begin{array}{c}
\pm i \chi'^1_-(- \varphi) \\ \chi'^1_+(- \varphi) \end{array} \right) =
R'_1(r) \left(\begin{array}{c}
\pm i \chi'^1_+(\varphi) \\ \chi'^1_-(\varphi) \end{array} \right) = 
\pm i \Psi'^1_\mp(\vec r).
\end{eqnarray}
Again, the corresponding eigenvalues of the 90 degrees rotation $O$ are 
$o = \pm i$, and hence, as for $\alpha \beta$ pairs, the symmetry is actually 
p-wave. Similarly, the two degenerate first excited states transform as
\begin{eqnarray}
^{D_i'}\Psi'^2_\pm(\vec r)&=&- R'_2(r) \left(\begin{array}{c}
\chi'^2_-(\varphi) \\ \pm \chi'^2_+(\varphi) \end{array} \right) = 
- \Psi'^2_\pm(\vec r), \nonumber \\
^O\Psi'^2_\pm(\vec r)&=&R'_2(r) \left(\begin{array}{c}
\pm \chi'^2_+(\varphi + \frac{\pi}{2}) \\ - \chi'^2_-(\varphi + \frac{\pi}{2}) 
\end{array} \right) = R'_2(r) \left(\begin{array}{c}
\mp \chi'^2_-(\varphi) \\ - \chi'^2_+(\varphi) \end{array} \right) = 
\mp \Psi'^2_\pm(\vec r), \nonumber \\
^R\Psi'^2_\pm(\vec r)&=&R'_2(r) \left(\begin{array}{c}
\pm \chi'^2_+(- \varphi) \\ \chi'^2_-(- \varphi) \end{array} \right) =
R'_2(r) \left(\begin{array}{c}
\pm \chi'^2_-(\varphi) \\ \chi'^2_+(\varphi) \end{array} \right) = 
\pm \Psi'^2_\pm(\vec r).
\end{eqnarray}
Again, the first excited states transform as s- or d-waves. The resulting 
probability distributions, which resemble $d_{xy}$ symmetry, are illustrated in 
figure 6 for the ground state (left panel) and the first excited state (right 
panel).
\begin{figure}[tb]
\begin{center}
\vspace{-0.3cm}
\epsfig{file=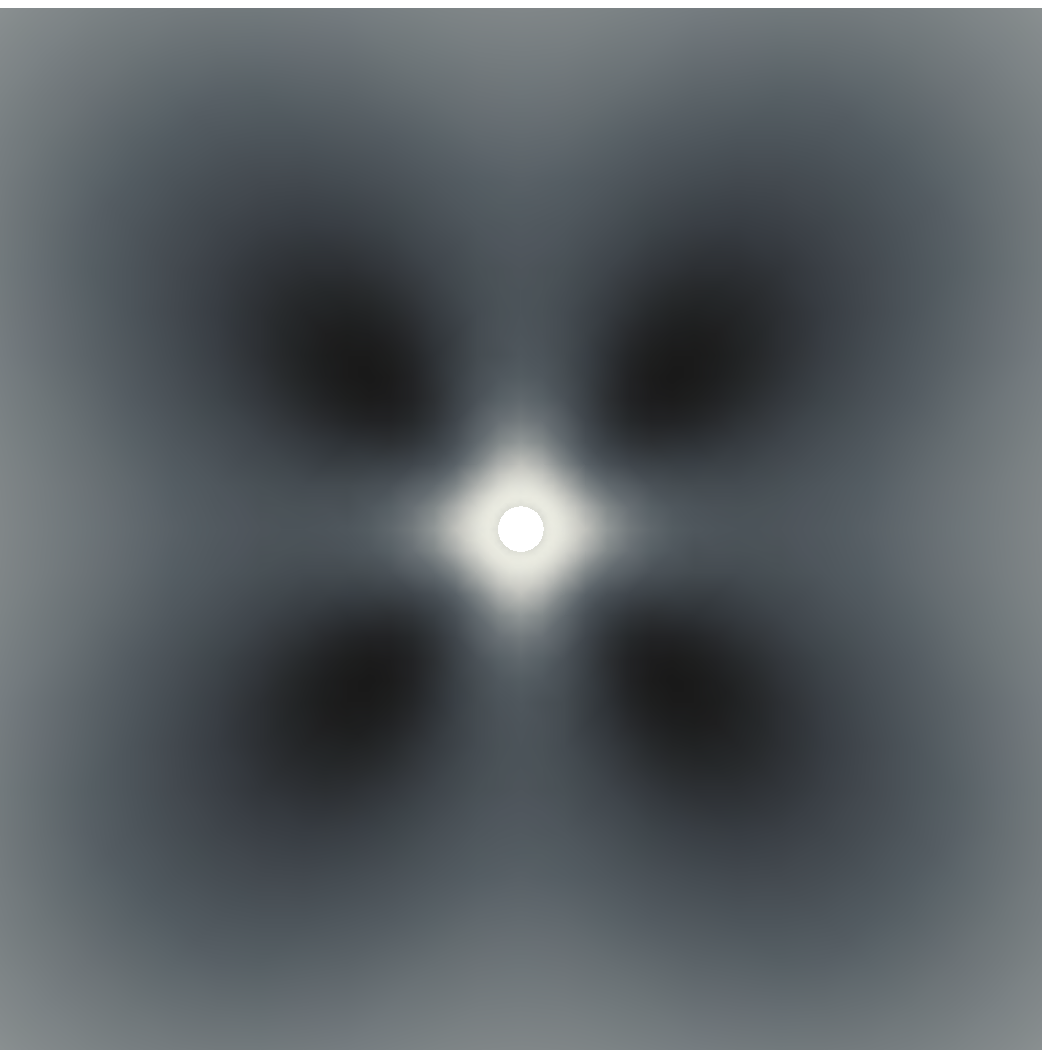,width=7.5cm}
\epsfig{file=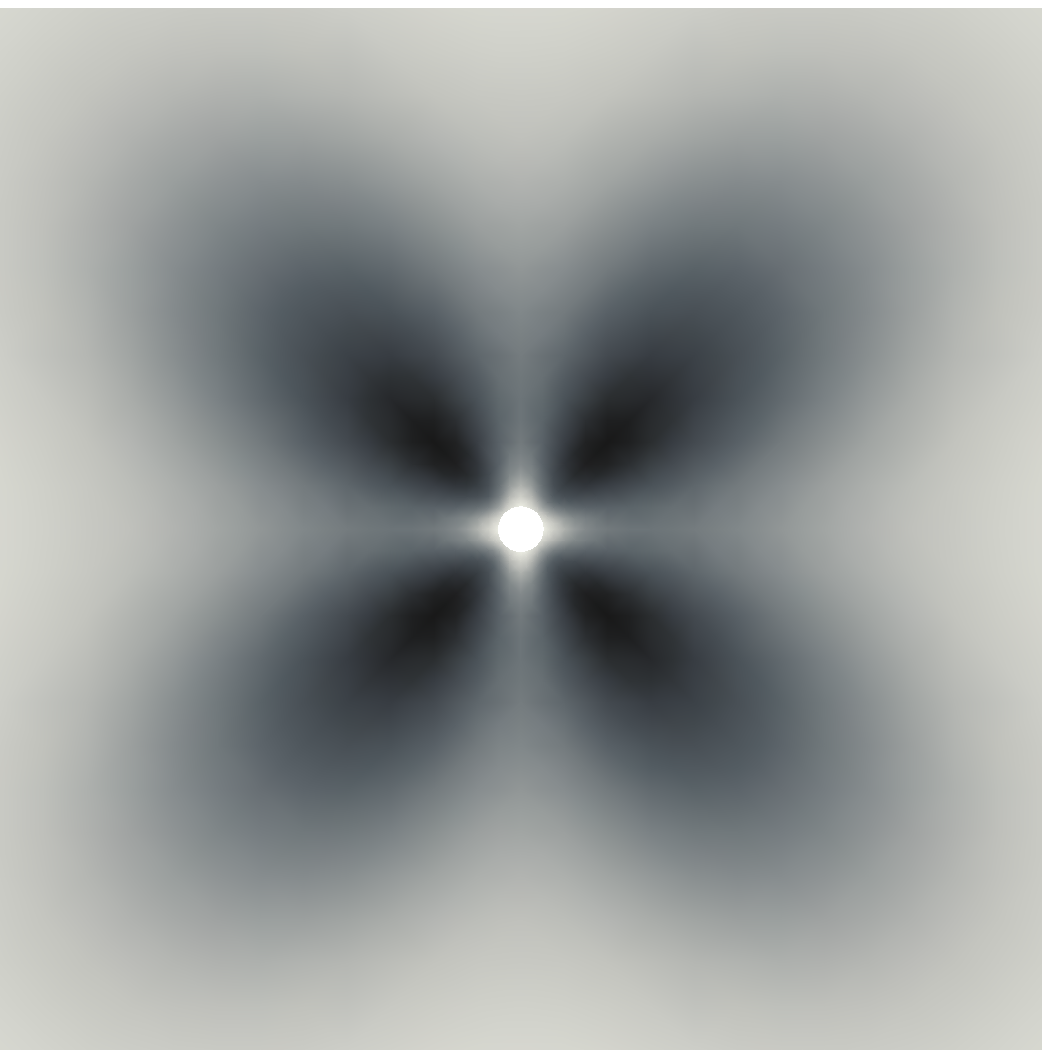,width=7.5cm}
\end{center}
\caption{\it Probability distribution for bound states of two holes with 
flavors $\alpha \alpha$ or $\beta \beta$, combined to an eigenstate of the 90
degrees rotation symmetry $O$. Left panel: the ground state with p-wave 
symmetry. Right panel: excited states with s- or d-wave symmetry, but with 
identical probability densities 
($M' \Lambda^2/4 \pi \rho_s = 1.25$, $r'_0 = a$).}
\end{figure}
Unlike for an $\alpha\beta$ pair, in the same flavor case the lowest energy
bound states localized on a Skyrmion have a different transformation behavior 
than the magnon-mediated two-hole bound states. 
 
\end{appendix}

\end{document}